\documentclass[accepted]{uai2023} 

\usepackage[american]{babel}

\usepackage{natbib} 
\usepackage{mathtools} 

\hypersetup{
colorlinks   = true, 
urlcolor     = blue, 
linkcolor    = blue, 
citecolor    = blue 
}

\usepackage{booktabs} 
\usepackage{tikz} 

\usepackage{url}            
\usepackage{booktabs}       
\usepackage{amsfonts}       
\usepackage{nicefrac}       
\usepackage{microtype}      
\usepackage{xcolor}         
\usepackage{algorithm}
\usepackage{algpseudocode}
\usepackage{float}

\usepackage{graphicx}
\usepackage{caption}
\usepackage{subcaption}

\usepackage{xspace}
\usepackage{amsthm}
\usepackage{amsmath}
\usepackage{enumitem}
\usepackage{booktabs}
\usepackage{multirow}
\usepackage{graphicx}
\usepackage[normalem]{ulem}
\useunder{\uline}{\ul}{}
\usepackage{float}
\usepackage{caption}
\usepackage{subcaption}
\usepackage{balance}
\usepackage[toc,page]{appendix}

\usepackage{amssymb}
\usepackage{amsmath}
\usepackage{amsfonts}

\DeclareMathOperator{\argmin}{\arg\min}
\DeclareMathOperator*{\minimize}{\text{minimize}}
\DeclareMathOperator*{\maximize}{\text{maximize}}

\title{Personalized Federated Domain Adaptation for Item-to-Item Recommendation}

\author[1,3]{Ziwei Fan\thanks{Corresponding Author.}}
\author[3]{Hao Ding}
\author[3]{Anoop Deoras}
\author[2]{Trong Nghia Hoang\thanks{Mentor. Corresponding Author. }}
\affil[1]{%
    University of Illinois Chicago\\
    Illinois, Chicago, USA
}
\affil[2]{%
    Washington State University\\
    Washington, Pullman, USA
}
\affil[3]{%
    AWS AI Lab\\
    California, Santa Clara, USA
  }
  
\begin{document}
\maketitle

\begin{abstract}
  Item-to-Item (I2I) recommendation is an important function in most recommendation systems, which generates replacement or complement suggestions for a particular item based on its semantic similarities to other cataloged items. Given that subsets of items in a recommendation system might be co-interacted with by the same set of customers, graph-based models, such as graph neural networks (GNNs), provide a natural framework to combine, ingest and extract valuable insights from such high-order relational interactions between cataloged items, as well as their metadata features, as has been shown in many recent studies. However, learning GNNs effectively for I2I requires ingesting a large amount of relational data, which might not always be available, especially in new, emerging market segments. To mitigate this data bottleneck, we postulate that recommendation patterns learned from existing mature market segments (with private data) could be adapted to build effective warm-start models for emerging ones. To achieve this, we propose and investigate a personalized federated modeling framework based on GNNs to summarize, assemble and adapt recommendation patterns across market segments with heterogeneous customer behaviors into effective local models. 
  Our key contribution is a personalized graph adaptation model that bridges the gap between recent literature on federated GNNs and (non-graph) personalized federated learning, which either does not optimize for the adaptability of the federated model or is restricted to local models with homogeneous parameterization, excluding GNNs with heterogeneous local graphs. 
\end{abstract}

\section{Introduction}
Item-to-Item (I2I) recommendation \citep{Li2019,Karypis2011} is a crucial product feature in most e-commerce businesses whose up-selling strategies depend on how well relevant items can be identified and curated into appealing suggestions to the customers. A key challenge to building such recommendation models is the need to compute item embeddings that not only encode their semantic features but also their high-order relations stemming from the heterogeneous preferential behaviors of customers, e.g. subsets of items in a recommendation system might be co-interacted with by similar subset of customers, indicating potential up-selling synergies among them.

To model such high-order relations, graph neural network (GNN) \citep{William18NIPS,KipfW17,GAT2018,DBLP:journals/corr/abs-1902-07153,DBLP:journals/corr/abs-1810-00826} was proposed and has been shown to be effective when applied to recommendation \citep{zhou2020graph}. However, building an effective GNN model for I2I recommendation often requires ingesting a large amount of annotated relational data, which is not always available in a new market segment, e.g., when a corporate decides to expand its business to a new user demographic. To mitigate this bottleneck, one approach is to adapt recommendation insights from other existing, more mature market segments to build a warm-start model. To achieve this while protecting the privacy of data, as mandated by strict data regulations such as the EU's GDPR \citep{DBLP:journals/corr/abs-1907-07498}, there have been recent developments that incorporate advances in federated learning \citep{McMahan17} into GNNs, resulting in a new modeling paradigm of federated GNNs \citep{DBLP:journals/corr/abs-2104-07145,DBLP:journals/corr/abs-2106-02743,DBLP:journals/corr/abs-2009-07351,wu2021fedgnn,zhang2021subgraph}.

Nonetheless, most of these prior approaches \citep{DBLP:journals/corr/abs-2104-07145} assume a central storage of the interaction graph or use GNN inductively where federated learning is enabled only on the feature propagation weights of the GNNs \citep{DBLP:journals/corr/abs-2106-02743,DBLP:journals/corr/abs-2009-07351}. Alternatively, more recent works also enable federated learning on predicting missing edges between local graphs \citep{zhang2021subgraph} or use encryption \citep{wu2021fedgnn} to communicate local graphs anonymously. However, such mechanisms either require sharing the private list of items that were interacted with across private markets \citep{zhang2021subgraph} or must have assumed no overlapping items across market segments \citep{wu2021fedgnn} because otherwise, the local encryption is not be able to recognize the same item that occurs in different market segments. Furthermore, all these works focus exclusively on learning a single GNN based on multiple decentralized data sources, implicitly assuming that the distributions or embedded structures of those sources are highly similar, e.g. co-interacted items across market segments were driven by similar customer preferential behaviors. This proves to be inaccurate across highly heterogeneous segments, leading to inferior performance of federated GNN approaches, as shown in (Section~\ref{sec:data}).

This is not surprising since such deficiencies of federated learning have been shown before in systems with highly heterogeneous clients \citep{DBLP:journals/corr/abs-2006-08848}, where the federated model is not necessarily the most adaptable model (on average) towards local data distributions of the clients. As such, personalized federated learning (FL) approaches \citep{DBLP:journals/corr/abs-2006-08848,DBLP:journals/corr/abs-1908-10400} have recently been devised to optimize for this most adaptable model, as well as its best adaptation towards local clients simultaneously, reporting successes in several applications. However, prior approaches in personalized federated learning are currently restricted to homogeneous model parameterization across clients, which excludes GNNs with heterogeneous graphs. Meanwhile, prior works in federated GNNs \citep{DBLP:journals/corr/abs-2104-07145,DBLP:journals/corr/abs-2106-02743,DBLP:journals/corr/abs-2009-07351,wu2021fedgnn,zhang2021subgraph}, as mentioned above, either do not enable federated learning on graphs, require sharing private item information or adopt non-differentiable techniques such as encryption to communicate and assemble graph information, which is not amenable to existing personalized FL approaches.

{\bf Key Contribution.} This motivates us to consider and explore potential distributed graph-based modeling techniques that bridge this gap between personalized federated learning and GNNs, as well as connect them to a new application of federated domain adaptation in I2I recommendation, which has been less explored in the context of GNNs. Our technical contributions include:


{\bf 1.} An adaptable graph summarization built on prior strategies in graph embedding \citep{kipf2016variational}, which encodes statistical structure that underlies the local market segment's observations of item interactions. The encoded structures do not expose item embeddings and can be communicated to a server, which combines them into a global structure optimized for high adaptability (Section~\ref{sec:graph}).

{\bf 2.} An adaptation of a recent (non-graph) personalized federated learning framework that incorporates the above graph summarization into its modeling mechanism, which enables personalized adaptation across local GNNs on both item feature and item-interaction structure levels (Section~\ref{sec:perFL}).

{\bf 3.} Extensive empirical studies on two large-scale cross-market item-to-item market datasets derived from the publicly available Amazon review dataset. Our studies first show that there are substantial heterogeneities in the item-interaction structures across market segments, which lead to inferior performance of vanilla federated GNN baselines in comparison to our proposed framework (Section~\ref{sec:exp}).

For interested readers, we also provide a detailed discussion on the broader topic of FL with GNNs in Appendix 10.

\section{Problem Definition}
\label{sec:def}


Let $\mathbb{C}$ denote the item catalogue and let $\mathbf{X} = \{\mathbf{x}_a\}_{a \in \mathbb{C}}$ denote the set of feature vectors $\mathbf{x}_a$ describing the attributes of each item $a$ in the catalogue $\mathbb{C}$. Then, let $\mathbb{O} = \{(\mathbf{x}_a^{(i)}, \mathbf{x}_{b^+}^{(i)}), (\mathbf{x}_a^{(i)}, \mathbf{x}_{b^-}^{(i)})\}_{i=1}^p$ denote a relationship dataset comprising $p$ positive and $p$ negative examples of the pairwise relationship. The item-to-item recommendation problem is then formulated as an embedding task which optimizes for an embedding function $\mathbf{z}_a = z(\mathbf{x}_a;\phi)$ parameterized by $\phi$ such that the inner product of item embeddings is most informative of the relationship between items via a logistic regressor,
\begin{eqnarray}
\label{eq:pred_eq}
r^{(i)}_{ab} &=& \sigma\Big(\theta_1 \cdot \left\langle\mathbf{z}_a^{(i)}, \mathbf{z}_b^{(i)}\right\rangle + \theta_2\Big) \ , \label{eq:2}
\end{eqnarray}
where $r^{(i)}_{ab} \in (0, 1)$ denotes whether the \emph{strength} of the relationship between items $a$ and $b$, and $\sigma(\cdot)$ is the sigmoid function. Both regressing parameters $(\theta_1, \theta_2)$ and the embedding function can be optimized via the averaged Bayesian Personalized Ranking (BPR) loss~\citep{RendleFGS09} via sigmoid function $\sigma(\cdot)$,
\begin{eqnarray}
\hspace{-12mm}\mathbf{L}\left(\theta, \phi\right) &=& -\frac{1}{p}\sum_{i=1}^p \Bigg(\log\sigma\left(r^{(i)}_{ab^+} - r^{(i)}_{ab^-}\right) \Bigg)  \ , 
\label{eq:3}
\end{eqnarray}
where $\theta = (\theta_1, \theta_2)$ and $\phi$ denotes the embedding parameterization, which is later detailed in Eq.~\eqref{eq:1b} of Section~\ref{sec:gnn}.

{\bf Key Challenges.} Despite being seemingly straightforward to optimize, the above formulation relies crucially on the characterization of the embedding function, which is non-trivial given that the relationship data is (1) decentralized across different market segments and cannot be centralized for model training; (2) not explicitly indicative of the complex high-order relationship between items stemming from the heterogeneous preferential behaviors of customers, which has been further occluded by their decentralized nature. The embedding function must therefore be robust against such heterogeneities and able to utilize all information for model training despite their private nature.

{\bf Solution Impact.} In practice, each of these challenges can in fact be addressed in isolation given the existing literature. For example, robust model training with heterogeneous and decentralized data can be enabled via personalized federated learning; while modeling the high-order relationship between items based on pairwise relational data can be achieved via graph-based models such as graph neural networks (GNNs). However, existing personalized federated learning has not been extended towards a graph-based model; and graph-based models are also not amenable to federated learning (FL) if the graph structure is not centralized, occluding full model visibility of the FL server. Our {\bf key contribution} here is therefore to {\bf combine these isolated solutions into a more coherent, robust solution that accommodates both personalized FL and GNN with decentralized graph data}, as detailed later in Section~\ref{sec:method}.

\section{Preliminaries} 
Our solution is developed based on key building blocks of graph neural network, variational graph auto-encoder and personalized federated learning which are reviewed below.

\subsection{Graph Neural Network}
\label{sec:gnn}
As mentioned above, we use the GNN model to encode the high-order relationship between items within each market segment into the item embedding function $\mathbf{z}_a = z(\mathbf{x}_a; \phi)$.
To ease the technical presentation, we will summarize the key idea of GNN using its most basic form as presented in the seminal work of \citep{kipf2017semisupervised}. We will then present its simplified graph convolution network (SGCN) formulation \citep{DBLP:journals/corr/abs-1902-07153}, which eases the computation cost and is adopted in our framework. 
For interested readers, there exists a larger literature on GNNs which is detailed in \citep{DBLP:journals/corr/abs-2104-07145} and \citep{zhou2020graph}.

To begin, a standard GNN is indexed by a graph $\mathbb{G} = (\mathbb{V}, \mathbb{E})$ with $n$ nodes which can be succinctly represented by an adjacency matrix $\mathbf{A}$. Each node $i$ in the graph is associated with a $d$-dimensional feature vector $\mathbf{x}_i \in \mathbb{R}^d$. For ease of notation, these feature vectors are often organized as rows of a matrix $\mathbf{X} \in \mathbb{R}^{n\times d}$, which is also treated as input to the GNN block. Given $\mathbf{X}$ and a fixed matrix $\mathbf{A}$, a GNN aims to produce a lower-dimensional embedding $\mathbf{z}_i \in \mathbb{R}^k$ for each node $i$ in the graph, which again can be succinctly organized as rows of a matrix $\mathbf{Z} \in\mathbb{R}^{n\times k}$. 


To avoid repeating the heavy matrix computation 
across multiple layers, we instead use the following practical approximation \citep{DBLP:journals/corr/abs-1902-07153},
\begin{eqnarray}
\hspace{-18mm}\mathbf{Z} &=& \sigma\Bigg(\Big(\mathbf{D}^{-\frac{1}{2}}\Big(\mathbf{A} + \mathbf{I}\Big)\mathbf{D}^{-\frac{1}{2}}\Big)^{m}\mathbf{X}\mathbf{W}\Bigg) \ , \label{eq:1b}
\end{eqnarray}
which collapses the recursion, allowing us to cache the computation of $\mathbf{D}^{-\frac{1}{2}}(\mathbf{A} + \mathbf{I})\mathbf{D}^{-\frac{1}{2}}$ in advance and consequently reducing both the forward and backward gradient computation cost of GNN. Eq.~\eqref{eq:1b} can be used as an effective characterization of the embedding function $z(\mathbf{x}_a; \phi)$ (with $\phi = (\mathbf{A}, \mathbf{W})$) in Section~\ref{sec:def}, which can be prepended to the logistic regressor featured in Eq.~\eqref{eq:1b} to enable end-to-end model training. However, this characterization in the distributed modeling context would expose the private local graph of each market segment and is therefore non-communicable or non-shareable. To work around this, we adopt graph embedding techniques such as the variational graph auto-encoder below to generate communicable summarization of local, private graphs. 

\subsection{Variational Graph Auto-Encoder}
\label{sec:vgae}
Let $\mathbb{G} = (\mathbb{V}, \mathbb{E})$ denote the relation graph that characterizes a certain relationship between items via its edges. Let $\mathbf{A}$ denote the corresponding adjacency matrix of $\mathbb{G}$. We view $\mathbf{A}$ as an observation drawn from a graph distribution conditioned item embeddings $\mathbf{Z}$, 
\begin{eqnarray}
\hspace{-2mm}p_{\theta}\left(\mathbf{A}|\mathbf{Z}\right) &=& \prod_{a=1}^n\prod_{b=1}^n \sigma\Big(\theta_1 \cdot \mathbf{z}_a^\top\mathbf{z}_b + \theta_2\Big)^{\mathbf{A}_{ab}}
\ \times\ \nonumber\\
\hspace{-2mm}&\ &\prod_{a=1}^n\prod_{b=1}^n \Big(1 - \sigma\Big(\theta_1 \cdot \mathbf{z}_a^\top\mathbf{z}_b + \theta_2\Big)\Big)^{1 - \mathbf{A}_{ab}} \label{eq:4}
\end{eqnarray}
where $\theta = (\theta_1, \theta_2)$ and $\mathbf{Z}$ is distributed a priori by a product of deep generative nets, $p_{\alpha}(\mathbf{Z}|\mathbf{X}) = \prod_{a=1}^n \mathbb{N}(\mathbf{z}_a; \mathbf{0},\mathrm{diag}[\alpha_a(\mathbf{X})])$ where $\alpha = \{\alpha_a\}_{a=1}^n$ and we assume $\mathbf{A} \perp \mathbf{X} \ |\  \mathbf{Z}$. We want to learn the optimal parameters $(\theta,\alpha)$ that maximizes
\begin{eqnarray}
\hspace{-24mm}\mathbf{G}\Big(\theta, \alpha\Big) &=& \log \mathbb{E}_{\mathbf{Z}\sim p_{\alpha}}\left[p_{\theta}\Big(\mathbf{A}|\mathbf{Z}\Big)\right] \ .\label{eq:4b}
\end{eqnarray}
Solving the optimization task in Eq.~\eqref{eq:4b} directly, however, appears intractable, even in the numeric sense as it is not clear how an unbiased stochastic gradient can be derived. To sidestep this issue, we learn a surrogate deep generative net that acts as the posterior of $\mathbf{Z}$ on observing $(\mathbf{X}, \mathbf{A})$,
\begin{eqnarray}
\hspace{-16mm} q_{\phi}\left(\mathbf{Z}|\mathbf{A},\mathbf{X}\right) &=& \prod_{a=1}^n\mathbb{N}\Big(\mathbf{z}_a | \mathbf{m}_a, \mathrm{diag}\left[\mathbf{v}_a\right]\Big) \ ,\label{eq:5}
\end{eqnarray}
where $\phi = (\phi_1, \phi_2)$, $\mathbf{m}_a$ is the $a^{\mathrm{th}}$ row of $\mathbf{M} = \mathrm{GNN}_{\phi_1}(\mathbf{X},\mathbf{A})$ and $\mathbf{v}_a$ is the $a^{\mathrm{th}}$ row of $\mathbf{V} = \mathrm{GNN}_{\phi_2}(\mathbf{X},\mathbf{A})$ where the two GNN blocks are parameterized separately with $\phi_1$ and $\phi_2$. This allows us to derive and optimize instead a lower-bound of the model evidence,
\begin{eqnarray}
\hspace{-1mm}\log p\left(\mathbf{A}|\mathbf{X}\right) &\geq& \mathbf{L}(\theta, \alpha, \phi) \ =\ \mathbb{E}_{q_{\phi}}\Big[\log p_{\theta}(\mathbf{A}|\mathbf{Z})\Big]\ -\ \nonumber\\
&&\mathbb{D}_{\mathrm{KL}}\Big(q_{\phi}\left(\mathbf{Z}|\mathbf{X},\mathbf{A}\right) \| p_{\alpha}(\mathbf{Z}|\mathbf{X})\Big)\ , \label{eq:6}
\end{eqnarray}
which is easier since it is expressed as expectation over $q_{\phi}$ whose parameterization is decoupled from $(\theta,\alpha)$, exposing a straight-forward unbiased stochastic gradient for those parameters. As for $\phi$, one can re-parameterize it using the trick in \citep{Kingma13} to expose a similar unbiased stochastic gradient. With this, $q_\phi$ is associated with the item embedding function $z(\mathbf{x}_a; \phi)$ (see Section~\ref{sec:def}), which can be optimized via Eq.~\eqref{eq:6} instead of Eq.~\eqref{eq:3}.


\subsection{Personalized Federated Learning}
\label{sec:PFL}
The increasingly decentralized nature of data generated in our digital society has stimulated the recent development of federated learning (FL) \citep{McMahan17}, which enables model training from multiple sources of private data without requiring them to leave their local sites. However, despite its advantage in protecting data privacy, FL is challenged by the statistical diversity across multiple data sources, which deviates from its implicit assumption that data distributions across sources are identical and independent. As has been shown in recent studies \citep{DBLP:journals/corr/abs-2006-08848,DBLP:journals/corr/abs-1908-10400}, ignoring this leads to federated models that do not generalize well. Intuitively, this happens because the federated model was not optimized for its adaptability.

To address this shortcoming, personalized federated learning \citep{DBLP:journals/corr/abs-2006-08848} was recently proposed by augmenting the training loss of FL into a bi-level optimization problem:
\begin{eqnarray}
\mathbf{w}_\ast \hspace{-3mm}&=&\hspace{-3mm} \argmin_{\mathbf{w}}\left\{\frac{1}{p}\sum_{i=1}^p \left(\hspace{-1mm}\ell_i\Big(\zeta_i(\mathbf{w})\Big) \hspace{-0.5mm}+\hspace{-0.5mm} \gamma\Big\|\zeta_i(\mathbf{w}) - \mathbf{w}\Big\|_2^2\right)\right\}\nonumber
\end{eqnarray}
where $\zeta_i(\mathbf{w})$ is given below:
\begin{eqnarray}
\hspace{-15mm}\zeta_i(\mathbf{w}) \hspace{-1mm}&=&\hspace{-1mm} \argmin_{\boldsymbol{\theta}}\left\{\ell_i\Big(\boldsymbol{\theta}\Big) + \gamma\Big\|\boldsymbol{\theta} - \mathbf{w}\Big\|_2^2\right\} \ ,\label{eq:7}
\end{eqnarray}
where $\ell_i(\theta)$ denote the training loss of task $i$ with respect to model parameter $\theta$. Simply put, Eq.~\eqref{eq:7} aims to find a base model $\mathbf{w}$ which can be adapted via the local fine-tuning function $\zeta_i(\mathbf{w})$. 
One particular choice of $\zeta_i(\mathbf{w})$ is detailed above where $\mathbf{w}$ is treated as a reference point for the fine-tuned model $\boldsymbol{\theta}$: $\boldsymbol{\theta}$ needs to reduce the training loss $\ell_i(\boldsymbol{\theta})$ while remaining sufficiently close to $\mathbf{w}$. This bi-level optimization task can then be solved effectively using the numerical approach described in \citep{DBLP:journals/corr/abs-2006-08848}. Alternatively, with a (somewhat) more artificial choice of $\gamma = 1/(2\alpha)$ and $\zeta_i(\mathbf{w}) = \argmin_{\theta}(\langle\nabla_{\mathbf{w}}\ell_i(\mathbf{w}), \boldsymbol{\theta} - \mathbf{w}\rangle + \gamma\|\boldsymbol{\theta} - \mathbf{w}\|_2^2)$, we have $\zeta_i(\mathbf{w}) = \mathbf{w} - \alpha\nabla_{\mathbf{w}}\ell_i(\mathbf{w})$, which is in closed-form. Thus, we can solve for the most adaptable $\mathbf{w}$ via 
\begin{eqnarray}
\mathbf{w} \leftarrow \mathbf{w} \ -\  \alpha \cdot \frac{1}{p}\sum_{i=1}^p\Big(\mathbf{I} - \alpha\nabla^2_{\mathbf{w}}\ell_i(\mathbf{w})\Big)\cdot \nabla_{\mathbf{w}'}\ell_i(\mathbf{w}') \label{eq:8}
\end{eqnarray}
where $\mathbf{w}' = \mathbf{w} - \alpha\nabla_{\mathbf{w}}\ell_i(\mathbf{w})$. This corresponds to the approach in \citep{DBLP:journals/corr/abs-1908-10400} which arises from a different derivation. Despite the artificial choices above, the resulting form of $\zeta_i(\mathbf{w})$ mimics the form of a one-step gradient update, indicating intuitively that $\mathbf{w}$ is being optimized with respect to its average fine-tuned performance after one step of gradient update, which is associated with its adaptability. Our notations are summarized in Table 1 of Appendix 1.

\begin{table*}[!h]
\centering
\begin{tabular}{|c|l|}
\hline
Notation & Definition \\ \hline
$n$ & The number of items \\ \hline
$\mathbf{A}\in\mathbb{R}^{n\times n}$ & The adjacency matrix of item-item affinity\\ \hline
$\mathbf{D}\in\mathbb{R}^{n\times n}$ & The degree matrix of $\mathbf{A}$\\ \hline
$\mathbf{X}\in\mathbb{R}^{n\times d}$ & Feature matrix of items \\ \hline
$\mathbf{Z}\in\mathbb{R}^{n\times k}$ & Low-dimensional item embeddings \\ \hline
$\hat{\mathbf{Z}}\in\mathbb{R}^{n\times k}$ & Approximated item embeddings \\ \hline
$\mathbf{O}$ & Item-item pairs dataset\\ \hline
$\theta = (\theta_1, \theta_2)$ & GNN I2I prediction layer parameters\\ \hline
$\alpha = \{\alpha_a\}_{a=1}^n$ & Parameterization of prior  $p_{\alpha}(\mathbf{Z}|\mathbf{X})$ in VGAE \\ \hline
$\mathbf{m}_a$ & Mean item embeddings from $\mathrm{GNN}_{\phi_1}(\mathbf{X},\mathbf{A})$ \\ \hline
$\mathbf{v}_a$ & Variance item embeddings from $\mathrm{GNN}_{\phi_2}(\mathbf{X},\mathbf{A})$ \\ \hline
$\phi = (\phi_1, \phi_2)$ & GNN layer parameters for $\mathrm{GNN}_{\phi_1}$ and $\mathrm{GNN}_{\phi_2}$ \\ \hline
$p$ & The number of market segment \\ \hline
$\mathbf{w}_\ast$ & The global GNN parameters \\ \hline
$\boldsymbol{\theta}$ & The local GNN parameters \\ \hline
$\ell_i(\boldsymbol{\theta})$ & Local training loss \\ \hline
$\mathbf{C}_{\boldsymbol{\kappa}^i}$ & The clusters assignment weights\\ \hline
$\boldsymbol{\kappa}^i$ & The cluster embedding of cluster $i$ \\ \hline
$\mathbb{P}_{\boldsymbol{\kappa}^i}$ & Differentiable clustering operator for cluster $i$ \\ \hline
$\phi_\ast$ & Global GNN parameters\\ \hline
$\xi_\ast$ & Global GNN summerization\\ \hline
$n_{\tau}$ & Number of global updates\\ \hline
$n_r$ & Number of local updates\\ \hline 
$\lambda_w$ & Local regularization weights on GNN parameters\\ \hline 
$\lambda_s$ & Local regularization weights on summarization\\ \hline
\end{tabular}
\caption{Notation Table}
\label{tab:notations}
\end{table*}


\section{Methodology}
\label{sec:method}
This section describes our approach to personalized federated domain adaptation via a communicable and adaptable graph summarization component that connects the above literature of personalized federated learning (Section~\ref{sec:PFL}) and GNNs (Section~\ref{sec:gnn}). This enables integration between personalized FL and GNNs, resulting in a robust federated domain adaptation model in Section~\ref{sec:perFL}.

\begin{figure*}[]
\centering
\includegraphics[width=0.85\textwidth]{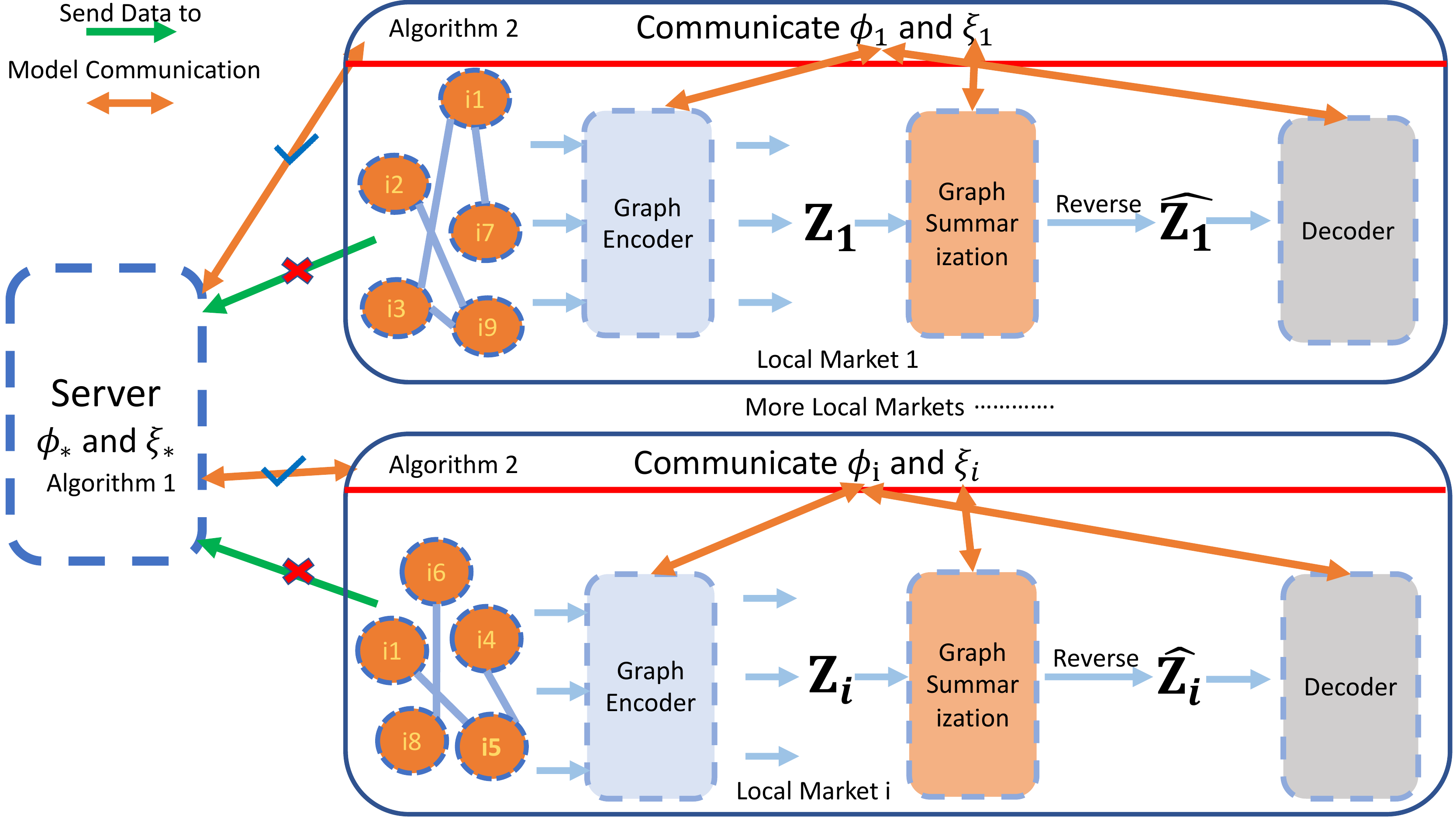}
\caption{Workflow Diagram. Each market consists of a graph encoder, which generates item embeddings $\bf{Z}$, the graph summarization process, the associated reverse operator, and a decoder for item-item relationship prediction. Each market shares and communicates $\phi$ (\textit{i.e.,} learnable parameters of the encoder) and $\xi$ (\textit{i.e.,} graph summarized structural information) with the server. Algorithm~\ref{alg:server} denotes the server optimization and Algorithm~\ref{alg:client} describes the client~(market) optimization. }
\label{fig:architecture}
\end{figure*}

\subsection{Adaptable Graph Summarization}
\label{sec:graph}
A key technical challenge in connecting personalized federated learning to GNNs stems from the facts that (a) personalized federated learning is mostly developed for scenarios with homogeneous local models, thus excluding GNNs whose graph structures are different across clients; and (b) GNNs can be extended to federated learning setting, but the resulting federated model is not optimized for its adaptability, resulting in inferior performance if client data are heterogeneous (see Section~\ref{sec:exp}). Furthermore, most of these federated GNN models \citep{DBLP:journals/corr/abs-2104-07145,DBLP:journals/corr/abs-2106-02743,DBLP:journals/corr/abs-2009-07351,zhang2021subgraph} do not enable federated learning on graph structure or require sharing private information.

As for the rest, maximizing the adaptability of the federated model following the above gradient-based optimization approach of personalized federated learning in Eq.~\eqref{eq:7} is challenged by the discreteness in their mechanisms, e.g., encryption \citep{wu2021fedgnn}, to communicate local graphs across clients. To sidestep such challenges, we propose the following two-phase structural encoding procedure:

{\bf Phase 1.} We adopt the variational graph auto-encoder in Section~\ref{sec:graph} to encode the structure of item interaction in each market segment into a differentiable representation, 
\begin{eqnarray}
\label{eq:phase1}
\hspace{-5mm}\maximize_{\theta_i,\alpha_i,\phi_i} &&\hspace{-2mm}\mathbf{L}_i(\theta_i, \alpha_i, \phi_i) \triangleq \mathbb{E}_{q_{\phi_i}}\Big[\log p_{\theta_i}(\mathbf{A}_i|\mathbf{Z})\Big] - \nonumber\\ \hspace{-2mm}&&\hspace{-2mm}\mathbb{D}_{\mathrm{KL}}\Big(q_{\phi_i}\left(\mathbf{Z}|\mathbf{X},\mathbf{A}_i\right) \| p_{\alpha_i}(\mathbf{Z}|\mathbf{X})\Big) \ , \label{eq:9}
\end{eqnarray}
where the subscript $i$ indexes the market segment. Once optimized, the local interaction graph $\mathbf{A}_i$ can be summarized via the encoder $q_{\phi_i}(\mathbf{Z} |\mathbf{X}, \mathbf{A}_i)$ described in Eq~\eqref{eq:5}.

{\bf Phase 2.} Embedding samples $\mathbf{Z}_i \sim q_{\phi_i}(\mathbf{Z}|\mathbf{X}, \mathbf{A}_i)$ can then be drawn from the (learned) local encoder $q_{\phi_i}$, which are compiled securely into local summaries $\xi_i = \mathbb{P}_{\boldsymbol{\kappa}^i}(\mathbf{Z}_i)$ where $\mathbb{P}_{\boldsymbol{\kappa}^i}$ is a differentiable operator parameterized by learnable parameter $\boldsymbol{\kappa}^i$ and adapted from a recent work on memory-based graph-level embedding \citep{DBLP:journals/corr/abs-2002-09518}. As such, the (local) differentiable summaries $\xi_i$ can be communicated on behalf of the less secure embedding $\mathbf{Z}_i$. The specific form of $\mathbb{P}_{\boldsymbol{\kappa}_i}$ is given via $\mathbb{P}_{\boldsymbol{\kappa}^i}(\mathbf{Z}_i) = \mathbf{C}^\top_{\boldsymbol{\kappa}^i}\mathbf{Z}_i$ where
\begin{eqnarray}
\label{eq:phase2_P_k}
\hspace{-11mm}\Big[\mathbf{C}_{\boldsymbol{\kappa}^i}\Big]_{ab} &\triangleq& \frac{\left(1 + \frac{1}{\tau}\left\|\mathbf{z}_a - \boldsymbol{\kappa}^i_b\right\|_2^2\right)^{-\frac{\tau + 1}{2}}}{\sum_{b'=1}^n \left(1 + \frac{1}{\tau}\left\|\mathbf{z}_a - \boldsymbol{\kappa}^i_{b'}\right\|_2^2\right)^{-\frac{\tau + 1}{2}} } \label{eq:10}
\end{eqnarray}
where $\mathbf{z}_a$ denotes the $a$-th row (or column) of $\mathbf{Z}_i$ and $\boldsymbol{\kappa}_b^i$ denotes the $b$-th row (or column) of $\boldsymbol{\kappa}^i$, with $\tau$ denoting a hyper-parameter to be empirically set to 1. Eq.~\eqref{eq:10} above implements a (soft) clustering of the item embeddings $\mathbf{z}_a \in \mathbf{Z}_i$ and aggregates items within a cluster based on membership weights $\mathbf{C}_{\boldsymbol{\kappa}^i}$. 

We then reverse the operator and approximately recover the item embedding $\widehat{\mathbf{Z}_i}$ via multiplying $\mathbb{P}_{\boldsymbol{\kappa}^i}(\mathbf{Z}_i)$ with the pseudo-inverse of $\mathbf{C}_{\boldsymbol{\kappa}^i}^\top$, which is  $(\mathbf{C}_{\boldsymbol{\kappa}^i}\mathbf{C}_{\boldsymbol{\kappa}^i}^\top)^{-1}\mathbf{C}_{\boldsymbol{\kappa}^i}$. The relationship prediction can then be conducted via Eq.~\eqref{eq:pred_eq} using $\widehat{\mathbf{Z}_i}$. This reverse operator makes $\xi_i = \mathbb{P}_{\boldsymbol{\kappa}^i}(\mathbf{Z}_i)$ viable summaries since once adapted, a local market can propagate the adaptation to the item embeddings via the inverse operator. This in turn enables the coordinating server to formulate a bi-level optimization task that searches for a global structure summary $\xi^\ast$ which is optimized for adaptability, following an adaptation of the personalized federated learning framework (Section~\ref{sec:perFL}).

{\bf Remark.} The above operator can be extended in both breadth and depth by combining ensemble and nested structures. For instance, we can organize $\mathbb{P}_{\boldsymbol{\kappa}^i}$ as a composition of multiple mappings, reflecting a hierarchical clustering structure. Each single mapping can be organized as a clustering ensemble, e.g. $\mathbf{C}_{\kappa^i} = \mathrm{softmax}(\mathrm{conv}[\mathbf{C}_{\kappa^i}^1, \ldots, \mathbf{C}_{\kappa^i}^q])$ where the $\mathrm{conv}$ operator is a $1 \times 1$ convolution on the concatenation of clustering structures \citep{DBLP:journals/corr/abs-2002-09518}.

\subsection{Personalized FL with GNN}
\label{sec:perFL}
As described in Section~\ref{sec:graph}, all market segments' graph summaries $\xi_i$ are the core of our domain adaptation approach and are generated based on high-level clustering representations of (local) item embeddings. Such summaries can be aggregated across market segment to pool insights and then further adapted to fit a local context. These summaries are also differentiable and can be deciphered back into a discrete interaction graph, making it a suitable medium for communication and adaptation. 

As such, we can succinctly characterize and demarcate local models for each market segment in terms of their segment-specific parameters $(\theta_i,\alpha_i)$ and segment-adaptable parameters $(\phi_i, \boldsymbol{\kappa}^i)$ which are in charge of graph decoding and encoding (i.e., summarizing), respectively. This demarcation is intuitive because $(\phi_i, \boldsymbol{\kappa}^i)$ parameterizes the process of summarizing the item embeddings across market segments, which need to encode adaptable information since the interaction activities across different segments happen on the same item catalog. Otherwise, $(\theta_i,\alpha_i)$ characterizes how the adapted item embeddings influence the interaction activities within a segment, which mostly depend on the user base of that segment. Hence, we view $(\theta_i,\alpha_i)$ as non-adaptable knowledge and will not enable personalized federated learning on them. Instead, we will enable personalized domain adaptation on $(\phi_i, \boldsymbol{\kappa}^i)$ by adapting the technique in Section~\ref{sec:graph}. This results in the following {\bf Phase 3} of training, which can be appended to the previous {\bf Phase 1} and {\bf Phase 2} of our proposed algorithm in Section~\ref{sec:graph}.

\begin{algorithm}
\caption{Server($n_\tau$, $n_r$, $\lambda_w$, $\lambda_s$) -- server in {\bf PF-GNN$+$}}
\label{alg:server}
\begin{algorithmic}[1]
\Require $n_{\tau}$, $n_r$, $\lambda_w$, $\lambda_s$ 
\Comment{$n_{\tau}$, $n_r$: no. global $\&$ local updates}\\\Comment{$\lambda_w$, $\lambda_s$ are adaptation moderating parameters}\\
initialize $\phi_\ast^{(0)}$ and $\xi_\ast^{(0)}$ \\\Comment{random feature $\&$ structure summaries}
\For{$t=1$ to $n_{\tau}$} \Comment{for each iteration}\\
    \hskip1.5em communicate $\xi_\ast^{(t-1)}$ and $\phi_\ast^{(t-1)}$ to all clients
    \For{all $i=1$ to $p$} \\
        \hskip3.0em $\phi_i^{(t)}, \xi_i^{(t)} \leftarrow \textbf{Client}\left(\phi_\ast^{(t-1)}, \xi_\ast^{(t-1)}, n_r\right)$\\\Comment{get personalized summaries from each client}
    \EndFor\\
    \hskip1.5em compute $\phi^{(t)}_\ast$ and $\xi_\ast^{(t)}$ via Eq.~\eqref{eq:13} \\
    \Comment{with $\zeta_i^1\left(\phi_\ast\right) = \phi_i^{(t)}$, $\zeta_i^2\left(\xi_\ast\right) = \xi_i^{(t)}$, $\lambda_w$ and $\lambda_s$}
\EndFor\\
\Return $\phi_\ast^{(n_{\tau})}$ and $\xi_\ast^{(n_\tau)}$ \hspace{-4mm}\Comment{return optimized summaries} 
\end{algorithmic}
\end{algorithm}

{\bf Phase 3.} First, we rewrite the training loss of each local market segment in Eq.~\eqref{eq:9} in the following succinct form, abstracting away the non-adaptable parameters,
\begin{eqnarray}
\label{eq:phase3_13}
\hspace{-11mm}\minimize_{\phi_i} \ \ \ell_i\left(\phi_i\right)  &\triangleq&
\min_{\theta_i, \alpha_i}\  \Big\{-\mathbf{L}_i\left(\theta_i, \alpha_i, \phi_i\right)\Big\} \label{eq:11}
\end{eqnarray}
where again $\mathbf{L}_i(\theta_i, \alpha_i, \phi_i)$ is defined in Eq.~\eqref{eq:9} above. Then, postulating that there exists a global adaptable parameter $\phi_\ast$ which can be fast adapted into any $\phi_i$ across different market segments, we can average their local losses and optimize for $\phi_\ast$ via a bi-level optimization task,
\begin{eqnarray}
\phi_\ast = \argmin_{\phi}\hspace{-2mm}&&\hspace{-6mm}\left\{\frac{1}{p}\sum_{i=1}^p \Bigg(\ell_i\Big(\zeta_i(\phi)\Big) + \lambda_w\Big\|\phi - \zeta_i(\phi)\Big\|_2^2\Bigg)\right\} \ \ \nonumber
\end{eqnarray}
where $\zeta_i(\phi)$ is given below:
\begin{eqnarray}
\hspace{-10mm}\zeta_i\left(\phi\right) &=& \argmin_{\phi_i} \left\{\ell_i\Big(\phi_i\Big) + \lambda_w\Big\|\phi - \phi_i\Big\|_2^2\right\} 
\label{eq:12}
\end{eqnarray}
where $\zeta_i(\phi)$ denotes the adaptation operator and $\lambda_w > 0$ is a tunable parameter that moderates the appropriate degree of adaptation. In addition, to enable personalized adaptation on graph summaries, we similarly postulate that there also exists a global summary $\xi_\ast$ which can be fast adapted into any summary $\xi^i$ of a local market segment -- see Eq.~\eqref{eq:10}. To optimize for $\xi_\ast$ and $\phi_\ast$, we augment Eq.~\eqref{eq:12} above to incorporate a structural regularization for $\xi_\ast$ via
\begin{eqnarray}
\label{eq:bilevel_optimization}
\hspace{-8mm}\Big(\phi_\ast, \xi_\ast\Big) \ \hspace{-3mm}&=&\hspace{-3mm} \ \underset{\phi,\xi}{\argmin}\ \frac{1}{p}\sum_{i=1}^p \Bigg\{\ell_i\Big(\zeta_i^1(\phi), \zeta_i^2(\xi)\Big) \nonumber\\
\hspace{-3mm}&+&\lambda_w\Big\|\phi - \zeta_i^1(\phi)\Big\|_2^2 
+\lambda_s \Big\|\xi - \zeta_i^2(\xi)\Big\|_2^2\Bigg\} \label{eq:13}
\end{eqnarray}
where the augmented local loss $\ell_i$ is defined as\vspace{-1.5mm}
\begin{eqnarray}
\label{eq:struct_optimize}
\hspace{-22mm}\ell_i(\phi_i, \xi_i) \ \hspace{-2mm}&=&\hspace{-2mm}\ \min_{\theta_i, \alpha_i, \boldsymbol{\kappa}^i} \Bigg\{-\mathbf{L}_i(\theta_i, \alpha_i, \phi_i) \nonumber\\
\hspace{-2mm}&+&\hspace{-2mm} \left\|\mathbf{A}_i - \mathbb{E}_{p_{\theta_i}}\Big[\mathbf{A} \ \Big|\ \mathbb{P}^{-1}_{\boldsymbol{\kappa}^i}\left(\xi_i\right)\Big]\right\|^2_{\mathrm{F}}\Bigg\} \ .\vspace{-3mm}
\label{eq:14}
\end{eqnarray}
Here, the adaptation operators are likewise defined as\vspace{-1.5mm}
\begin{eqnarray}
\hspace{-10mm}\Big(\zeta_i^1(\phi), \zeta_i^2(\xi)\Big) \ \hspace{-2mm}&=&\hspace{-2mm}\ \argmin_{\phi_i, \xi_i} \Bigg\{\ell_i(\phi_i, \xi_i)  \nonumber\\
\hspace{-2mm}&+&\hspace{-2mm} \lambda_w \Big\|\phi - \phi_i\Big\|_2^2
+ \lambda_s \Big\|\xi_i - \xi\Big\|_2^2\Bigg\} \ , \vspace{-3mm}\label{eq:15}
\end{eqnarray}
with $\lambda_w > 0$ and $\lambda_s > 0$ denote parameters that moderate the adaptation of $\phi$ and $\xi$ appropriately. In Eq.~\eqref{eq:14}, the added regularizer is meant to enforce that the adapted summary $\xi_i$ and the summarizing operator $\mathbb{P}_{\boldsymbol{\kappa}^i}$ can reconstruct the local interaction graph. Eq.~\eqref{eq:14} completes our proposed federated domain adaptation framework (Algorithms~\ref{alg:server} and~\ref{alg:client}) for cross-market item-to-item recommendation, whose workflow is summarized below.

\subsection{Server-Client Workflow}
\label{sec:g_l_optimization}
Our proposed federated domain adaptation proceeds in multiple iterations, as detailed in Algorithm~\ref{alg:server}. At each iteration, the server communicates the current global graph summaries to the clients (see Algorithm~\ref{alg:server}, line $4$). Each client  runs Algorithm~\ref{alg:client} to compute and return the corresponding personalized summaries to the server (see Algorithm~\ref{alg:client}, line $6$). Once all personalized summaries have been received, the server updates the global summaries via Eq.~\eqref{eq:13} (see Algorithm~\ref{alg:server}, line 10) and starts the next iteration. 

\subsection{Secure Computation}
\label{sec:g_l_optimization}
This section discusses potential privacy threats that might impact our work and suggests safe practice to avoid those unseen threats. First, although we are (to the best of our knowledge) not aware of any existing works that would enable such threats in context relevant to ours, we do not rule out unforeseen cases in which communicating graph representation might (arguably) be linked to new security threats (e.g., graph content might be reverse-engineered to expose sensitive customer information). To minimize exposure to such threats, we believe a safe practice is to incorporate existing privacy-preserving approaches to secure the communicating process. For example, off-the-shelves secure aggregation method such as~\citep{bonawitz2019federated} can be readily applied to our work. 

This help clients securely share the output of any processing function (e.g., solution of Eq.~(\ref{eq:bilevel_optimization})) of their local data with the server, allowing to learn the function aggregation without being able to infer additional information about the local data~\citep{kairouz2021advances}. This is possible since our proposed FL framework follows the same setting of existing FL methods~\citep{DBLP:journals/corr/abs-2006-08848, fallah2020personalized, McMahan17, li2021ditto}, which are compatible with such generic-purpose privacy-preserving methods.


\begin{algorithm}
\caption{Client($\phi_\ast, \xi_\ast$, $n_r$) -- client $i$ in {\bf PF-GNN$+$}}
\label{alg:client}
\begin{algorithmic}[1]
\Require $\phi_\ast$, $\xi_\ast$, $n_r$ \Comment{global summaries, no. of iterations}\\
initialize $\phi_i^{(0)}$, $\boldsymbol{\kappa}^i_{(0)}$ and compute $\xi_i^{(0)}$ via Eq.~\eqref{eq:10}\\  \Comment{local summaries for client $i$} 
\For{$t=1$ to $n_r$} \\
    \hskip1.5em solve for $\theta_i$ and $\alpha_i$ via Eq.~\eqref{eq:14} \\
    \Comment{fixing $\boldsymbol{\kappa}^i =\boldsymbol{\kappa}^i_{(t-1)}$, $\phi_i = \phi_i^{(t-1)}$ and $\xi_i = \xi_i^{(t-1)}$}\\
    \hskip1.5em solve for $\boldsymbol{\kappa}^i_{(t)}$ $\&$ $\phi_i^{(t)}$ via Eq.~\eqref{eq:15} \\
    \hskip1.5em setting $\xi_i^{(t)}$ as function of $\boldsymbol{\kappa}^i_{(t)}$ via Eq.~\eqref{eq:10} \\\Comment{$\xi_i^{(t)} = \mathbb{P}_{\kappa^i_{(t)}}(\mathbf{Z}_i)$ where $\mathbf{Z}_i \sim q_{\phi_i}(\mathbf{Z} \mid \mathbf{X},\mathbf{A}_i)$}
\EndFor\\
\Return $\phi_i^{(n_r)}$, $\xi_i^{(n_r)}$ \Comment{feature $\&$ structure summaries}
\end{algorithmic}
\end{algorithm}
\vspace{-3mm}

\section{Experiments}
\label{sec:exp}
We compare the performance of the proposed framework against those of several baselines (Section~\ref{sec:baseline}) on the cross-market recommendation dataset
\citep{DBLP:journals/corr/abs-2109-05929} (Section~\ref{sec:data}). Our experimental setup is detailed in Section~\ref{sec:setup}. 

\subsection{Dataset}
\label{sec:data}
The cross-market dataset features a comprehensive collection of user-item and item-item interactions, as well as item metadata across multiple market segments and domains on the same item catalog. The dataset spans multiple product categories and international market segments. Our empirical studies focus on items from the \emph{Electronic} and \emph{Home and Kitchen} domains, whose data are scattered across $12$ and $11$ international market segments, respectively. We filtered out market segments with fewer than 100 interactions. Due to limited space, the results of \emph{Home and Kitchen} are deferred to Appendix 5. These include Arabia, China, Australia, Japan, France, Spain, Germany, United Kingdom, India, Mexico, Canada, and America. The data statistics of Electronics domain are reported in Table 2 of Appendix 3.

\begin{table*}[h!]
	\centering
	\begin{sc}
	\begin{small}
		\begin{tabular}{|l|l|l|l|l|l|}
			\hline
			{\bf Arabia} & \hspace{-0.5mm}{\bf China} & \hspace{-0.5mm}{\bf Australia} & \hspace{-0.5mm}{\bf Japan} & \hspace{-0.5mm}{\bf France} & \hspace{-0.5mm}{\bf Spain}\hspace{-1mm}\\
			\hline
			$328$ / $6440$ & \hspace{-0.5mm}$1303$ / $2087$ & \hspace{-0.5mm}$2390$ / $4834$ & \hspace{-0.5mm}$4003$ / $41861$ & \hspace{-0.5mm}$6068$ / $1451380$ & \hspace{-0.5mm}$6572$ / $109166$\hspace{-1mm}\\
			\hline
			{\bf Germany} & {\bf UK} & {\bf India} & {\bf Mexico} & {\bf Canada} & {\bf United-States}\hspace{-1mm}\\
			\hline
			$7507$ / $159154$ & \hspace{-0.5mm}$10329$ / $441033$ & \hspace{-0.5mm}$6574$ / $23869$ & \hspace{-0.5mm}$8507$ / $139783$ & \hspace{-0.5mm}$18604$ / $400825$ & \hspace{-0.5mm}$35939$ / $2048177$\hspace{-1mm}\\
			\hline
		\end{tabular}
	\end{small}
	\end{sc}
	\caption{Data statistics (no. active items / no. unique interaction pairs) across different market segments. The numbers of active items and unique item-item interaction pairs among them are reported.}\vspace{-2mm}
	\label{tab:1}
\end{table*}


\subsection{Market Heterogeneity Analysis}
\label{sec:hetero}
Due to the different distribution of these interactions across market segments, the locally induced (latent) item embeddings and interaction structures are substantially heterogeneous. In particular, we compute the locally induced item embeddings via optimizing Eq.~\eqref{eq:9} for each market segment. Cosine similarities between observed pairs of interacting items (e.g., items reported to be frequently bought together) across all segment are computed and then partitioned into discrete bins. Each segment induces a categorical distribution over bins. The feature heterogeneity between two markets is set to be the Jensen-Shannon divergence between their induced distributions. This is visualized in Fig. 4a of Appendix 6 which shows moderate heterogeneity\footnote{Due to limited space, we decided to defer all figures of this section to Appendix 6}.

However, the heterogeneity becomes significantly more pronounced as we look into the interaction structure. Following the approach described in \citep{DBLP:journals/corr/abs-1805-11921}, we sample random walks along the edges of local item-interaction graphs. The sampled walks across all market segments are partitioned into clusters, enabling us to represent a segment in terms of a categorical distribution over a common space of random walks. The difference between two market segments can then be computed as the divergence between two corresponding distributions over random walks. This is visually reported in Fig.~4b of Appendix 6. As a result, this high degree of heterogeneity has rendered the naive transfer of a pre-trained model from one segment to another ineffective, as shown in Fig. 4c. This motivates us to consider a personalized federated learning solution to this problem where recommendation insights across segments are harnessed, exchanged, and communicated concurrently, resulting in better recommendations (see Fig.~\ref{fig:MRR_Electronics}).

\begin{figure*}[!ht]
     \centering
     \begin{subfigure}[b]{0.3\textwidth}
         \centering
         \includegraphics[width=1\textwidth]{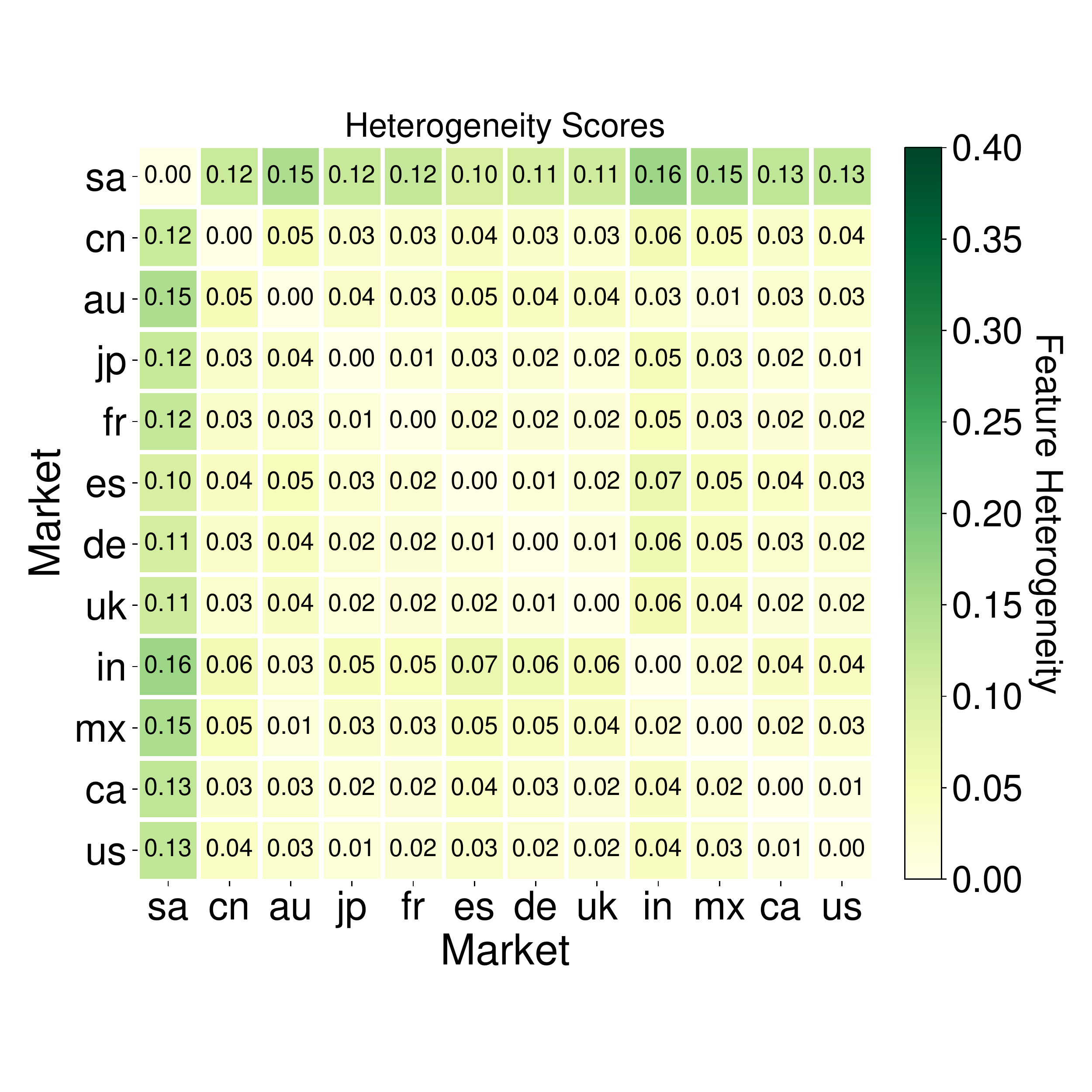}\vspace{-4mm}
         \caption{}\vspace{-3mm}
         \label{fig:hetero_feat_elect}
     \end{subfigure}\hfill
     \begin{subfigure}[b]{0.3\textwidth}
         \centering
         \includegraphics[width=1\textwidth]{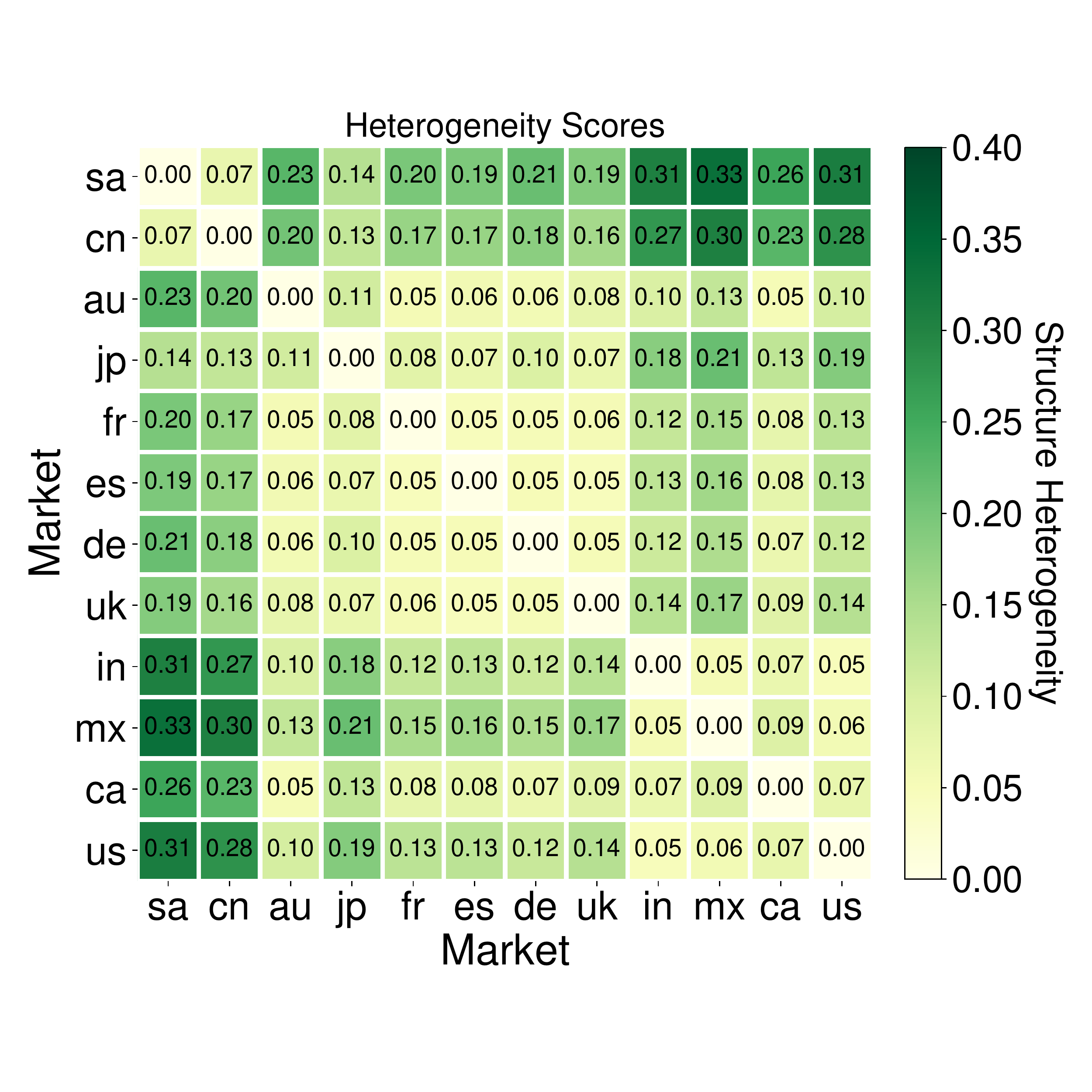}\vspace{-4mm}
         \caption{}\vspace{-3mm}
         \label{fig:hetero_struct_elect}
     \end{subfigure}\hfill
     \begin{subfigure}[b]{0.3\textwidth}
         \centering
         \includegraphics[width=1\textwidth]{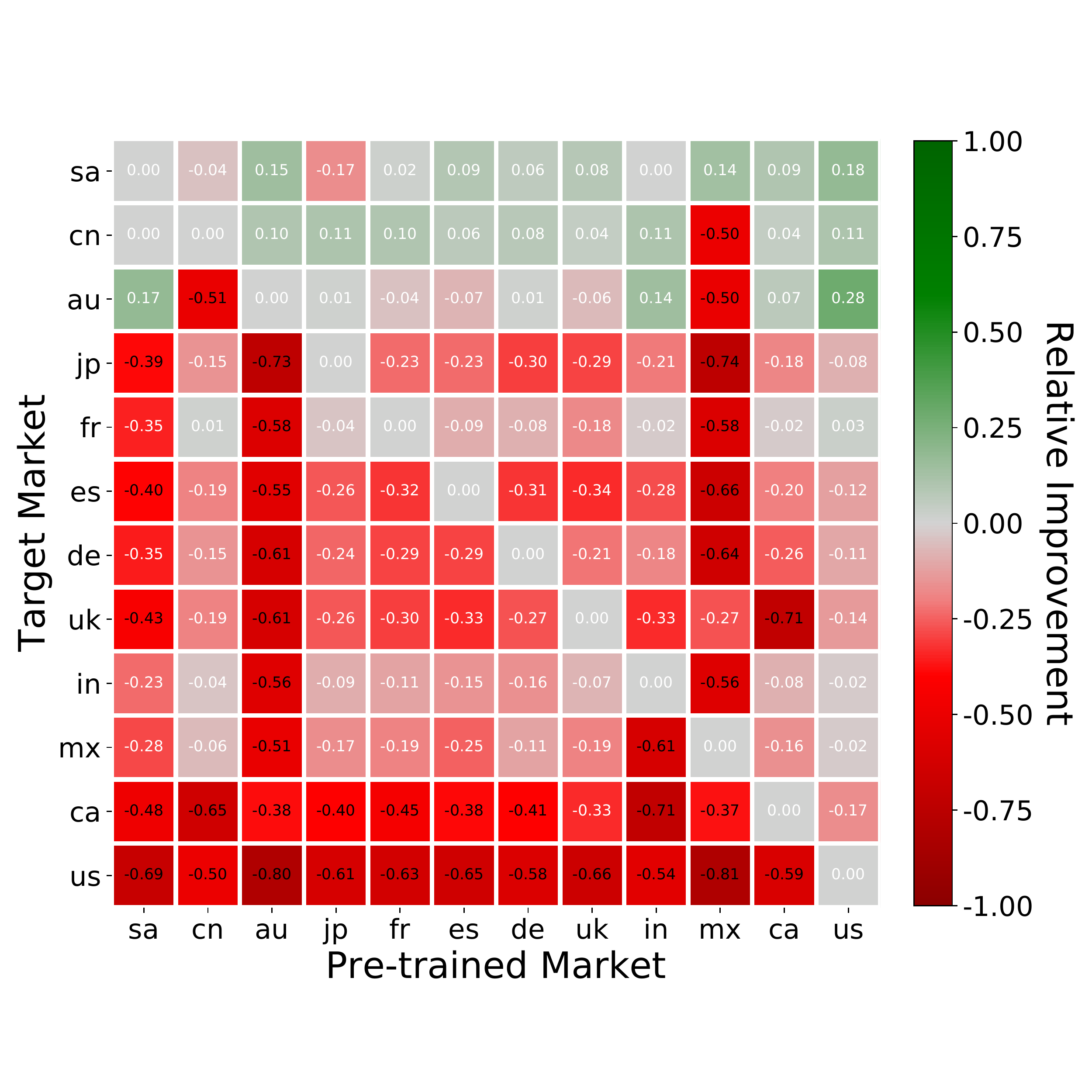}\vspace{-4mm}
         \caption{}\vspace{-3mm}
         \label{fig:indsgcn_compare_elect}
     \end{subfigure}
\caption{Plots of {\bf (a)} item embedding heterogenity across markets; {\bf (b)} item interaction heterogenity between across market segments; and {\bf (c)} negative effect of naive transfer of a pre-trained recommendation model on one segment to another (best view with color). The calculation of these heterogenity and (negative) naive transfer scores are detailed in Section~\ref{sec:hetero}.}\vspace{-2mm}
\label{fig:extra}
\end{figure*}

\subsection{Evaluation Metric}
\label{sec:metric}
To evaluate the quality of related-item recommendation for a queried item $\mathbf{x}$, we use two standard metrics: (a) the mean reciprocal rank (MRR); and (b) the normalized discounted cumulative gains (NDCG). Due to limited space, we defer the specification of these two metrics to the appendix.

\subsection{Experiment Setup and Evaluation}
\label{sec:setup}

{\bf Item Feature.} For each item in the catalog, we have additional information, such as its title and text description. We use a pre-trained BERT model\footnote{\url{https://huggingface.co/transformers/v3.3.1/pretrained_models.html}} to encode the concatenated title and description text of an item into a continuous feature space. This results in a $768$-dimensional feature vector per input node, which forms the feature matrix $\mathbf{X}$ of the GNN.

{\bf Item-Item Interaction.} In our dataset, we have access to customer access logs for each market segment that record events when a customer browses an item and also views another item within the same browsing session. This qualifies as an \emph{Also-Viewed} event. Additionally, we also have access to events when customers purchase both items in a session, which are categorized as \emph{Also-Bought} events. Both of these are used to describe item-item edges that denote their observed interaction. We reserve $90\%$ of such item pairs per segment for training purposes, leaving the remaining $10\%$ as local tests. The detailed hyper-parameters selections are in Appendix 7. To evaluate the quality of top-$N$ related-item recommendation for an item $\mathbf{x}$, we use two standard ranking metrics: (a) the mean reciprocal rank (MRR@N); and (b) the normalized discounted cumulative gains (NDCG@N).


\begin{figure*}[!ht]
     \centering
     \begin{subfigure}[b]{0.25\textwidth}
         \centering
         \includegraphics[width=1\textwidth]{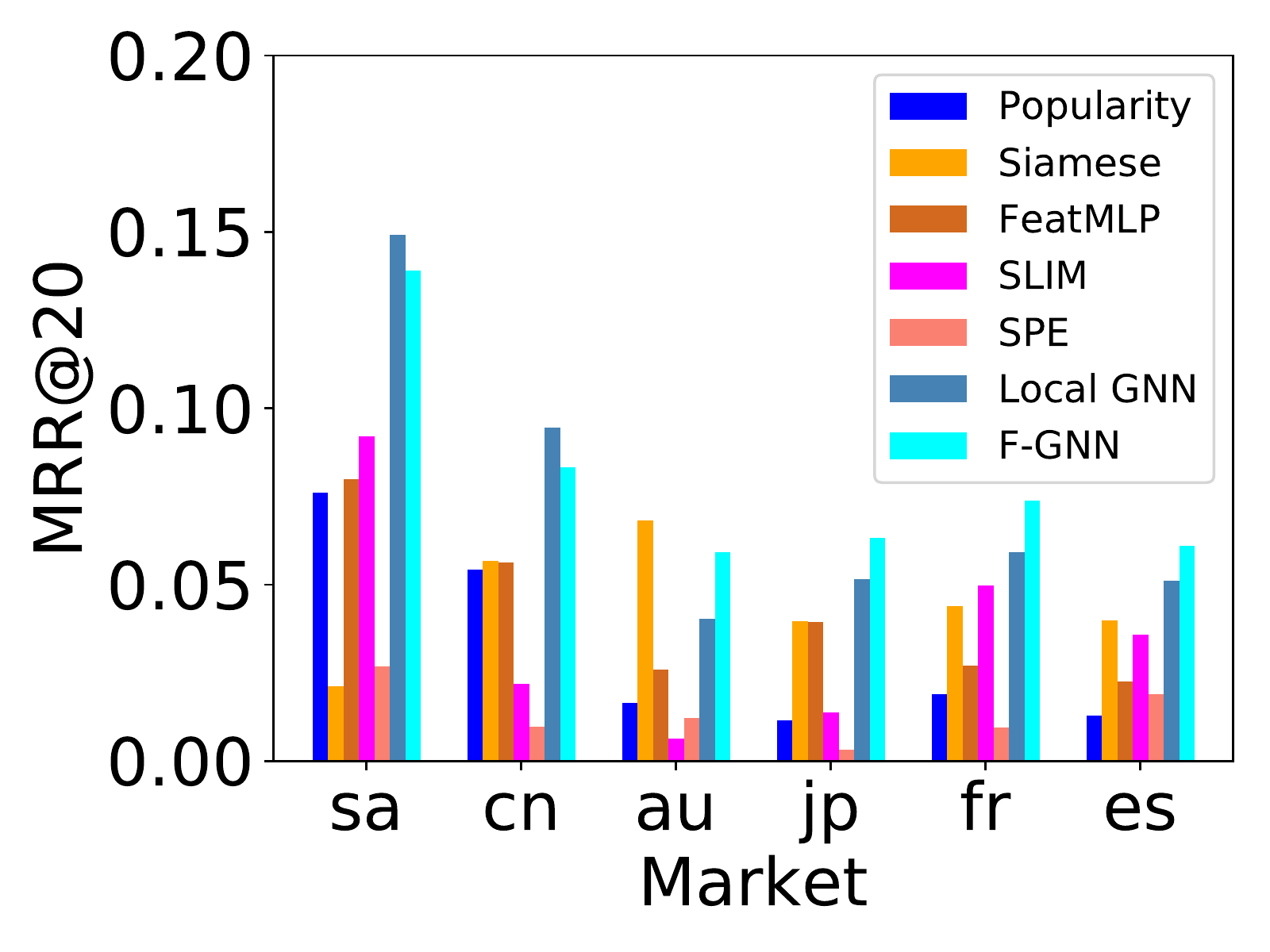}\vspace{-2mm}
         \caption{}\vspace{-1mm}
         \label{fig:low_re_baselines_mrr}
     \end{subfigure}\hfill
     \begin{subfigure}[b]{0.25\textwidth}
         \centering
         \includegraphics[width=1\textwidth]{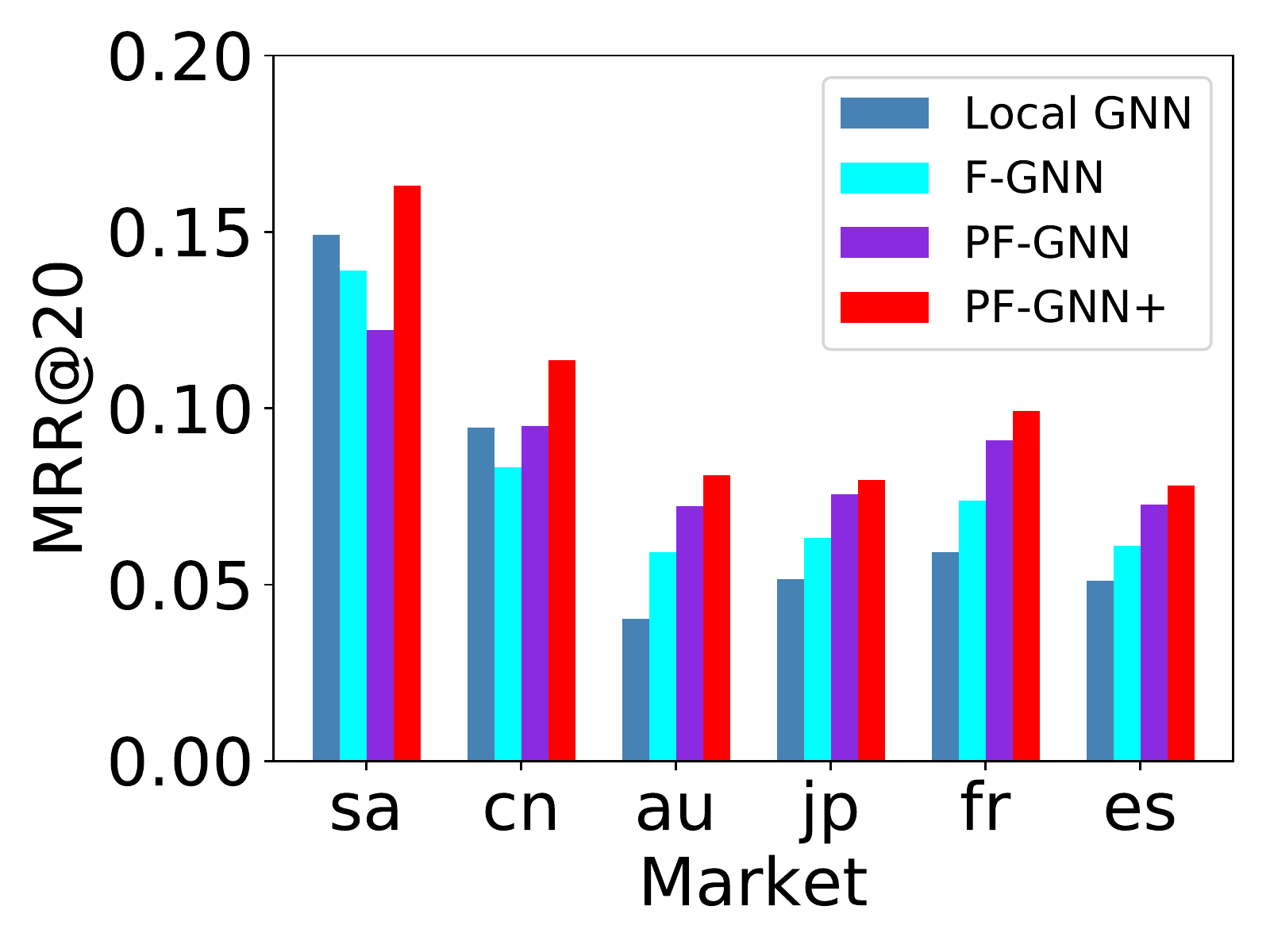}\vspace{-2mm}
         \caption{}\vspace{-1mm}
         \label{fig:mrr_baby}
     \end{subfigure}\hfill
     \begin{subfigure}[b]{0.25\textwidth}
         \centering
         \includegraphics[width=1\textwidth]{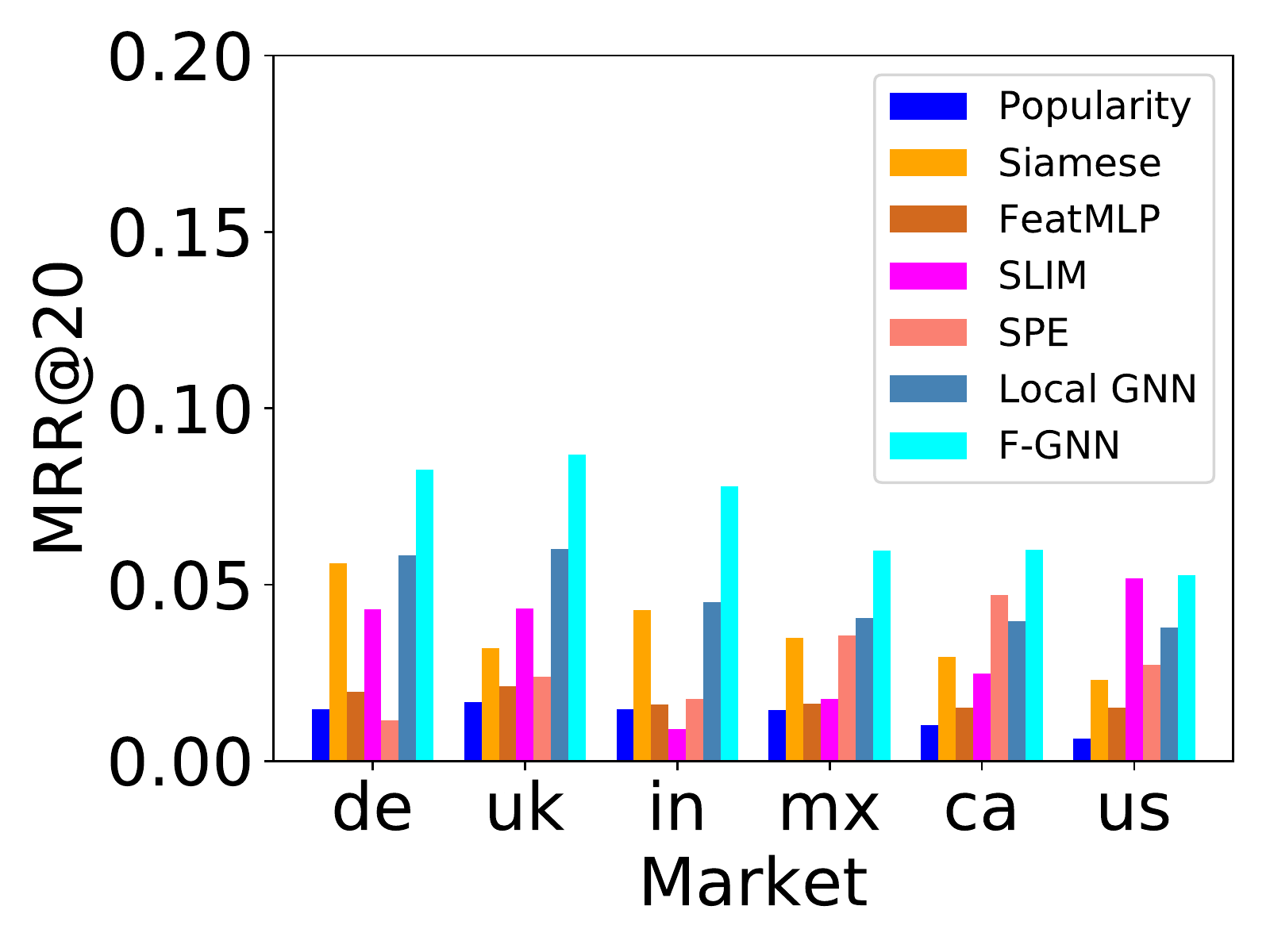}\vspace{-2mm}
         \caption{}\vspace{-1mm}
         \label{fig:mrr_tools}
     \end{subfigure}\hfill
     \begin{subfigure}[b]{0.25\textwidth}
         \centering
         \includegraphics[width=1\textwidth]{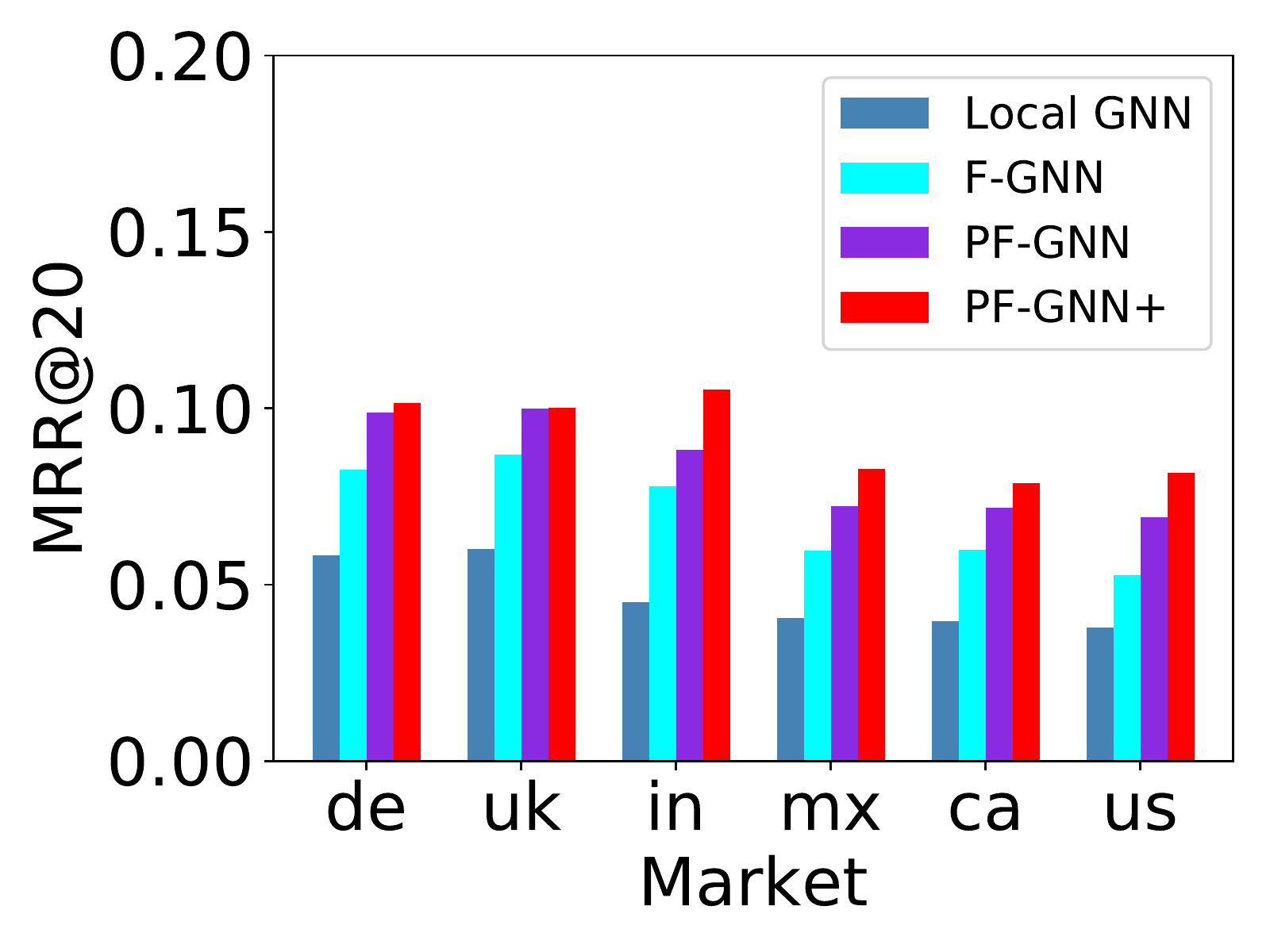}\vspace{-2mm}
         \caption{}\vspace{-1mm}
         \label{fig:mrr_music}
     \end{subfigure}\hfill\\

     \begin{subfigure}[b]{0.25\textwidth}
         \centering
         \includegraphics[width=1\textwidth]{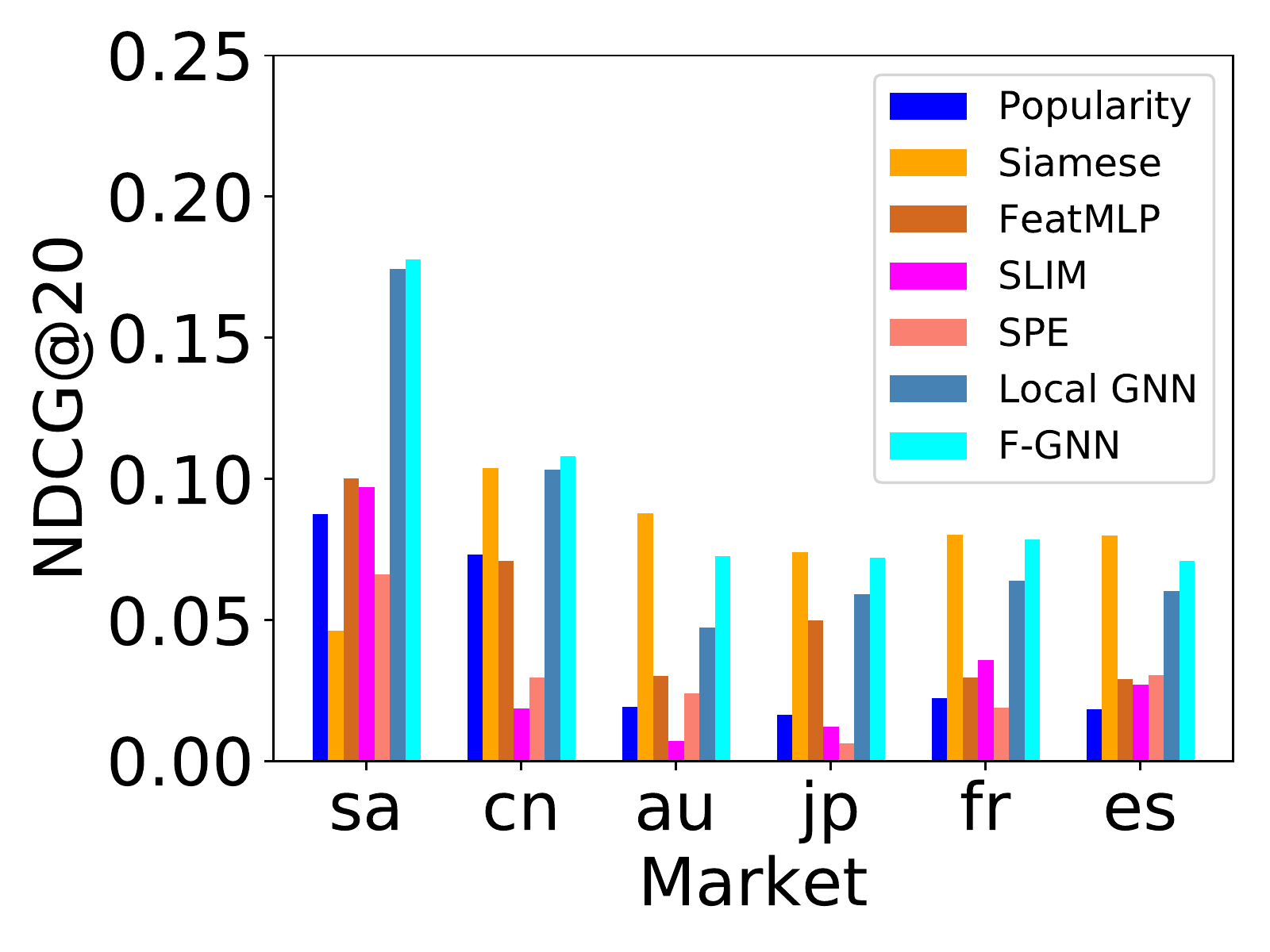}\vspace{-2mm}
         \caption{}\vspace{-2mm}
         \label{fig:mrr_toy}
     \end{subfigure}\hfill
     \begin{subfigure}[b]{0.25\textwidth}
         \centering
         \includegraphics[width=1\textwidth]{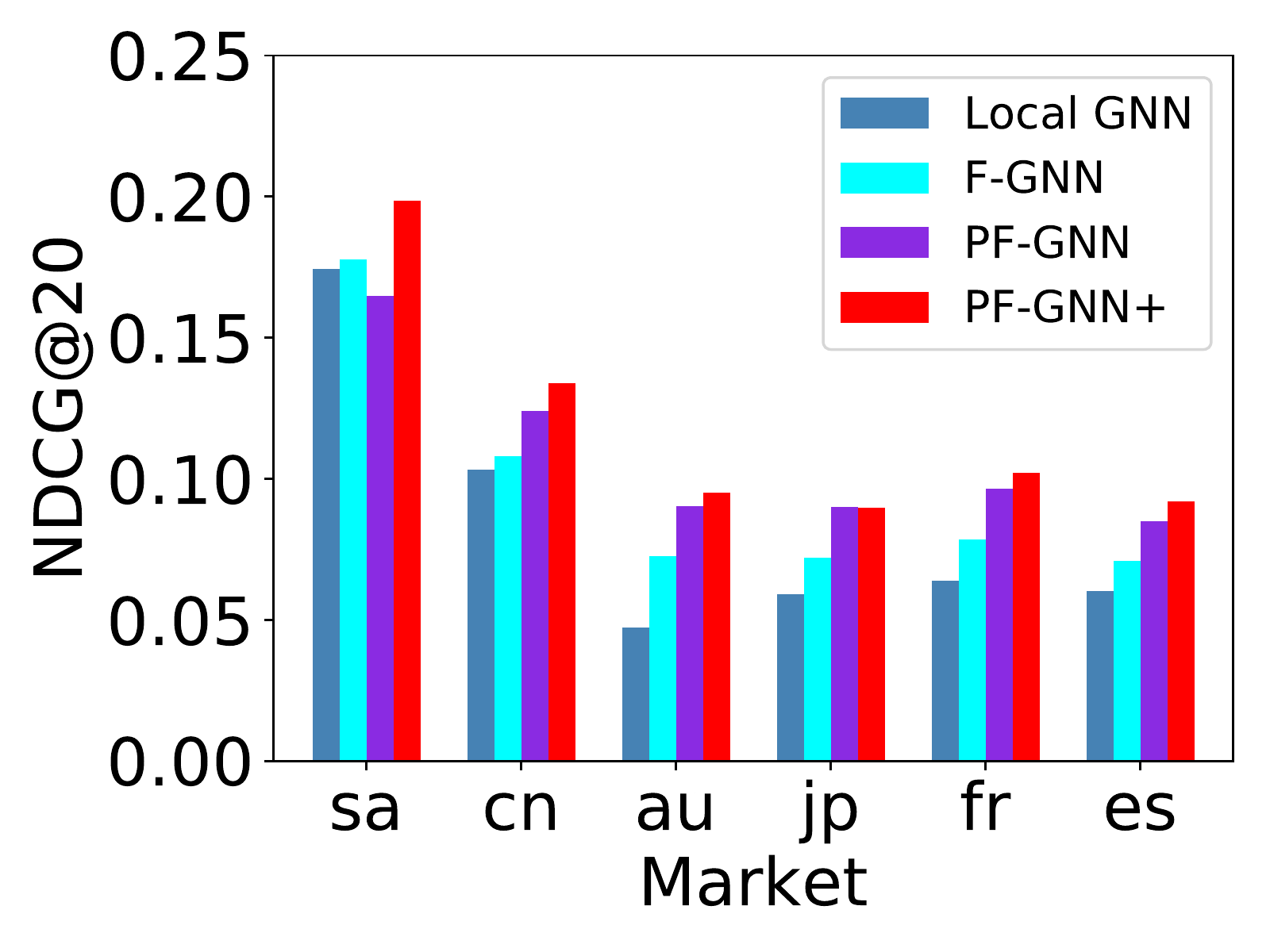}\vspace{-2mm}
         \caption{}\vspace{-2mm}
         \label{fig:mrr_baby}
     \end{subfigure}\hfill
     \begin{subfigure}[b]{0.25\textwidth}
         \centering
         \includegraphics[width=1\textwidth]{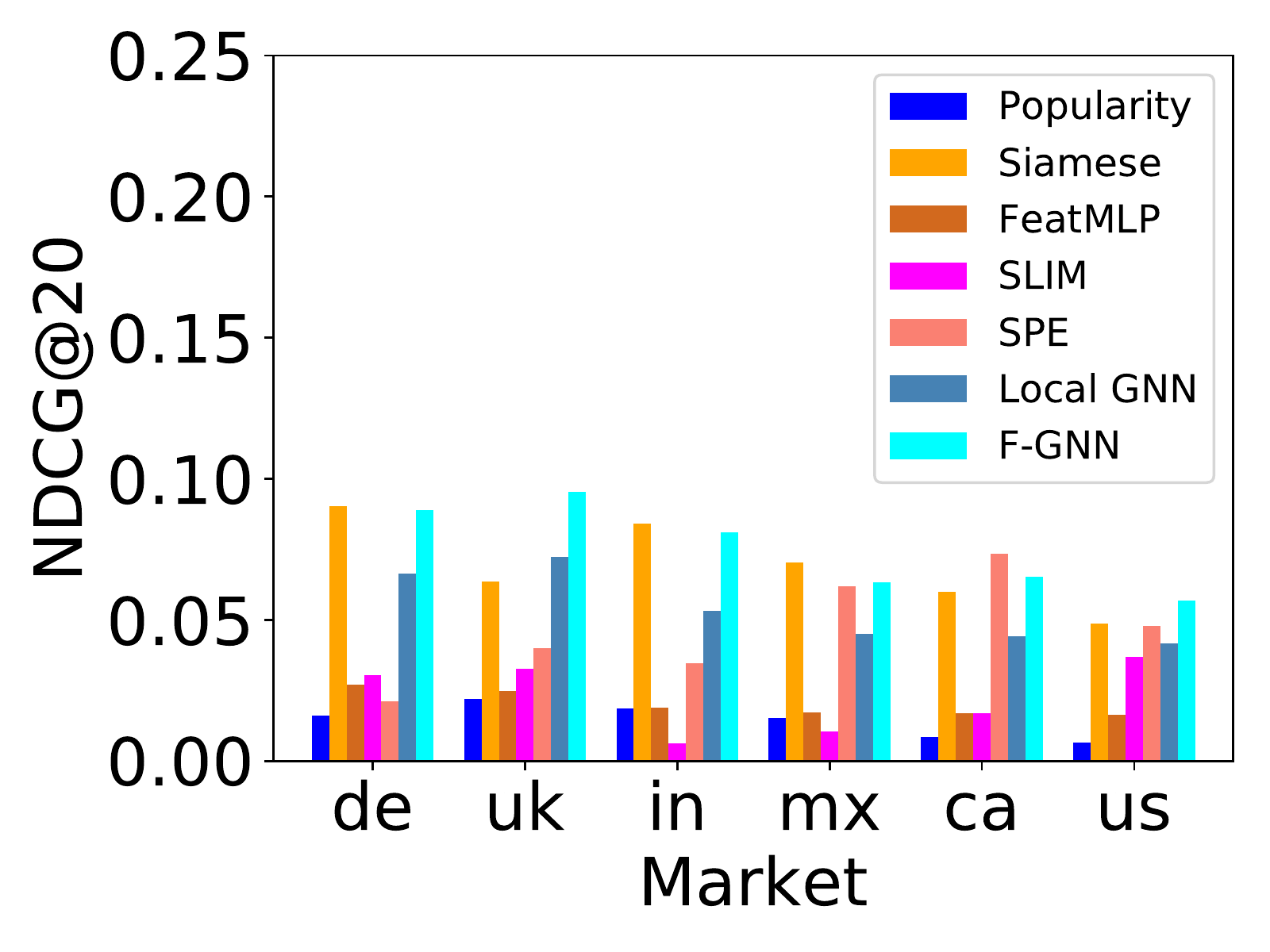}\vspace{-2mm}
         \caption{}\vspace{-2mm}
         \label{fig:mrr_tools}
     \end{subfigure}\hfill
     \begin{subfigure}[b]{0.25\textwidth}
         \centering
         \includegraphics[width=1\textwidth]{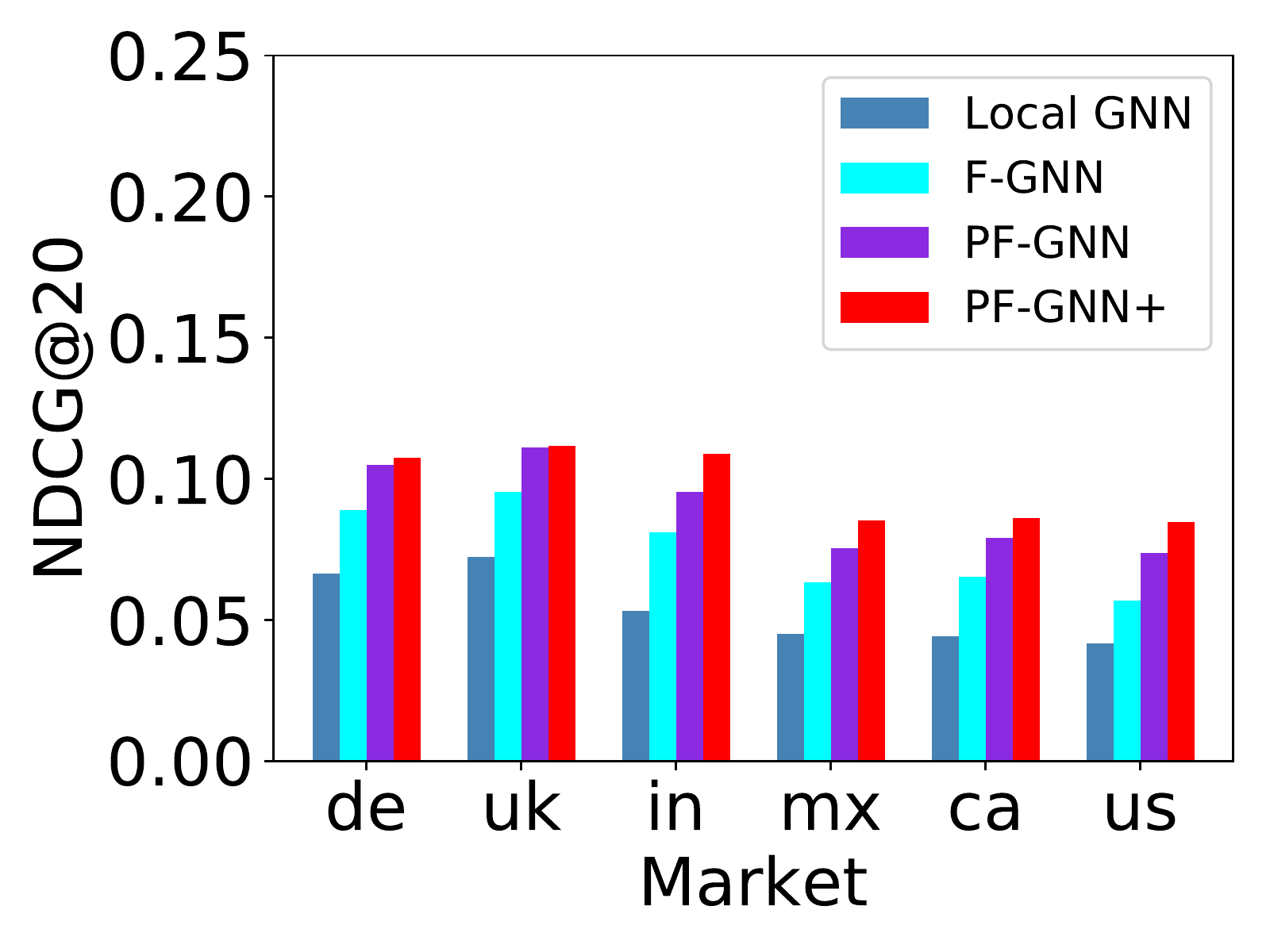}\vspace{-2mm}
         \caption{}\vspace{-2mm}
         \label{fig:mrr_music}
     \end{subfigure}
     \vspace{-6mm}
        \caption{Comparison of the {\bf MRR@20} and {\bf NDCG@20} recommendation metric between our personalized federated domain adaptation algorithms, including {\bf PF-GNN} and {\bf PF-GNN$+$}, and other baselines across different markets. To avoid cluttering plots, we split the performance report of all baselines into smaller plots. On the top row, we report the {\bf MRR@20} of the baselines across the market segments. The first two plots on the top row collectively report the {\bf MRR@20} performance of all participating algorithms in the first $6$ segments while the next two plots report for the remaining $6$ segments. For ease of comparison, we put the (same) performance bars of {\bf F-GNN} and {\bf Local GNN} baselines in both plots as a frame of reference for cross-plot comparison. We adopt the same plotting scheme for {\bf NDCG@20} in the bottom row. }\vspace{-3mm}
    \label{fig:MRR_Electronics}
\end{figure*}

\subsection{Hyper-parameter Configuration and Model Parameterization}
All of our experiments were conducted on a compute machine with $8$ V$100$ GPUs. For all GNN baselines, the GNN is parameterized with $3$ layers of Simplified Graph Convolution Network \cite{DBLP:journals/corr/abs-1902-07153} which map from an item's $768$-dimensional feature vector to a $128$-dimensional representation embedding vector. To boost performance, we also perform grid search for important parameters of the models such as the learning rate which varies within $\{0.1, 0.01, 0.001\}$; the feature aggregation adaptation parameter $\lambda_w$ in Eq.~\eqref{eq:13} within $\{0.1, 1, 10, 100, 1000\}$ which controls the importance of adapting feature aggregation via personalizing $\phi$; and the structure adaptation parameter $\lambda_s$ within $\{0.001, 0.01, 0.1, 1, 10\}$ which moderates the relative importance of item-item interaction structure adaptation via $\xi$ -- see last term in Eq.~\eqref{eq:13}. The best parameter configurations are selected based on their performance of US market. In particular, the best configuration of {\bf PF-GNN} is specified with $0.1$ for learning rate and $1000$ for $\lambda_w$. For {\bf PF-GNN$+$} which additionally involves the structure adaptation moderator $\lambda_s$, the best configuration is specified with the same learning rate but with different choices of $\lambda_w = 1$ and $\lambda_s = 0.01$.

\subsection{Baselines}
\label{sec:baseline}
The performance of our proposed algorithms are evaluated and compared against two GNN-driven baselines, which include: (a) {\bf Local GNN} which is a locally trained GNN \citep{DBLP:journals/corr/abs-1902-07153} following the formulation in Eq.~\eqref{eq:1b}\footnote{We note that this simplified parameterization of {\bf Local GNN} is highly similar to LightGCN~\citep{he2020lightgcn} which is reported to outperform NGCF~\citep{wang2019neural} with simpler formulation.}; 
(b) {\bf F-GNN} which is a federated GNN model as proposed in \citep{wu2021fedgnn}. In addition, we include to the set of baselines several other specific I2I recommendation methods, which include: (c) {\bf SLIM}~\citep{Karypis2011} which takes the user-item interaction matrix as input and generates the item-item correlations as item embeddings; 
(d) {\bf SPE}~\citep{Li2019} which proposes a semi-parametric approach to learn item embeddings; and 
(e) {\bf Siamese} which is a contrastive learning baseline for I2I recommendation recently used in~\citep{hoang2022learning}. 

We also compare our methods with two other vanilla baselines: (f) {\bf FeatMLP} that adds one MLP layer after the item features $\mathbf{X}$ to generate item embeddings, where $\mathbf{Z}_i=\mathbf{X}\mathbf{W}_i$ for each market $i$; and
\noindent(g) {\bf Popularity} which is a heuristic (non-learnable) algorithm that simply recommends the most popular items in the catalogue. Finally, two variants of our proposed framework are used in the above empirical evaluation: (a) {\bf PF-GNN} which enables personalized FL only on feature encoding parameters $\{\phi_i\}_{i=1}^p$ -- Eq.~\eqref{eq:10}; and (b) {\bf PF-GNN$+$} which further enables personalized FL for the structure encoding parameters $\{\boldsymbol{\kappa}^i\}_{i=1}^p$ -- Eq.~\eqref{eq:13}.\vspace{-0mm}

\subsection{Result Observations}
\label{sec:result_observations}
Results in the \emph{Electronics} category are reported in Fig.~\ref{fig:MRR_Electronics}. Repeated results in the same \emph{Electronics} with standard deviations are shown in Fig. 2 of Supplement Section 4. We include results of \emph{Home and Kitchen} category in Fig. 3 of Supplement Section 5. We have following observations:

{\bf 1.} The results show that among the participating algorithms, {\bf PF-GNN$+$} performs most robustly and produces the best performance in all $12$ market segments. Our other variant {\bf PF-GNN} produces the second-best performance, which is substantially better than the other methods in $11/12$ markets. Our averaged results over multiple runs reported in Fig. 2 of Supplement Section 4 further confirms the improved performance of {\bf PF-GNN$+$} and {\bf PF-GNN} over baselines.

{\bf 2.} {\bf PF-GNN$+$} consistently outperforms {\bf PF-GNN}. The difference between {\bf PF-GNN$+$} and {\bf PF-GNN} is that {\bf PF-GNN} only exchanges $\phi$ while {\bf PF-GNN$+$} also exchanges the graph structural information summary $\xi$. This demonstrates the necessity of graph summarization and its communication across market segments. 

{\bf 3.} {\bf F-GNN} is the best among all baselines (not including the variants of our proposed algorithm) in most cases. These results, therefore, support our hypotheses earlier that (a) a federated learning solution is necessary to enable robust domain adaptation; (b) federated GNNs, however are less effective when local data distributions are substantially heterogeneous as shown in Fig. 4; and (c) optimizing explicitly for the adaptability of the federated model often produced superior performance, which is demonstrated in both cases: with and without structure adaptation.

{\bf 4.} {\bf Siamese}, {\bf SPE}, and {\bf Local GNN} generally achieves better performance than other non-federated baselines. Among these, {\bf Local GNN} achieves more significant improvements in low-resource markets, which indicates the advantage of utilizing high-order information extracted from GNN for I2I recommendation. In larger markets, however, {\bf Siamese} and {\bf SPE} appear to perform better than {\bf Local GNN}, suggesting that for market segments with a large amount of interactions, local sub-graph is insufficient to encode the high-order interaction between items accurately. Local models, therefore, perform better by ignoring the graph structure and instead learning the contrastive information from the examples of pairwise item relationships, which are much more abundantly available.

\begin{figure}[]
\centering
     \begin{subfigure}[b]{0.235\textwidth}
         \centering
         \includegraphics[width=1\textwidth]{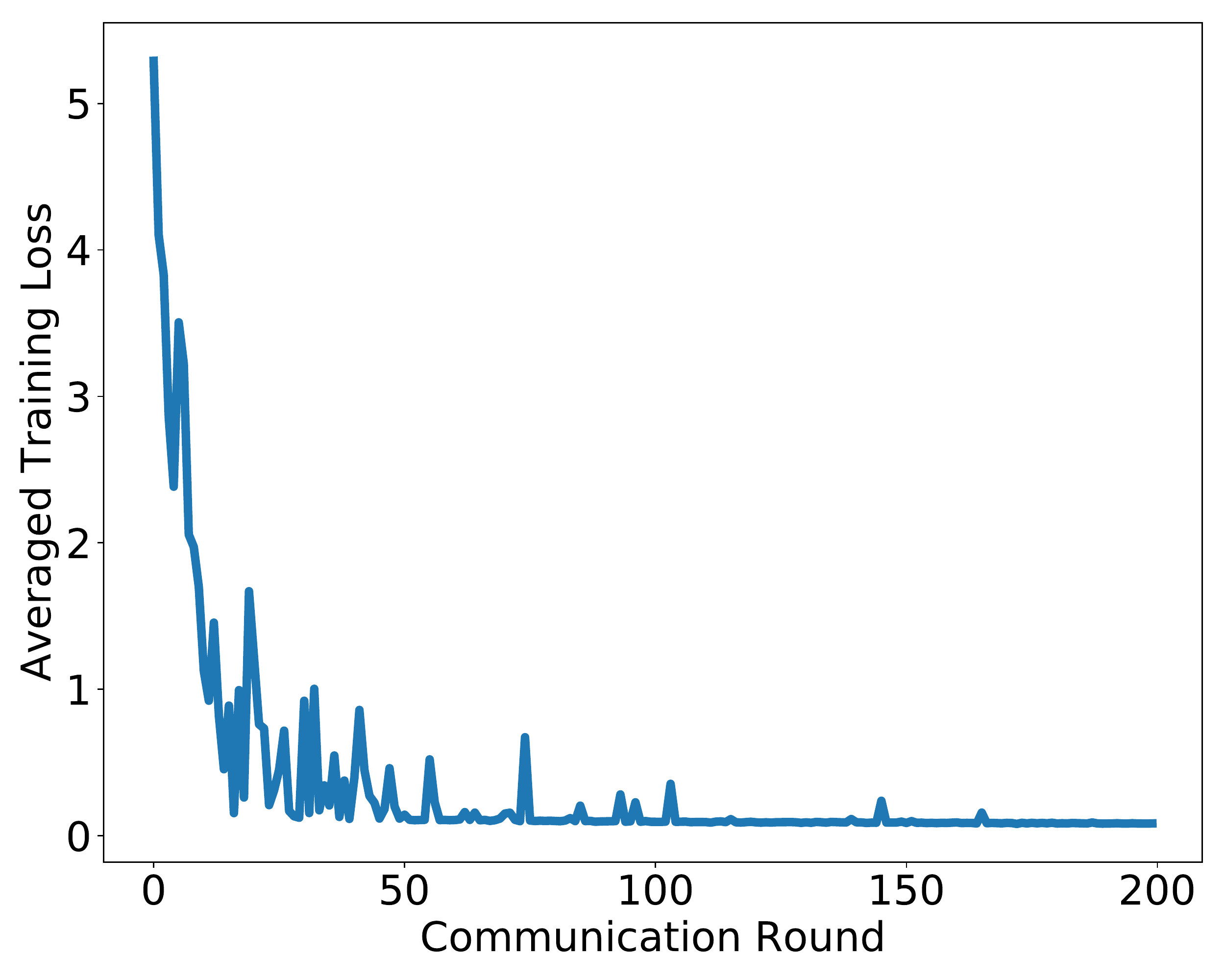}
         \caption{{\bf PF-GNN}}
         \label{fig:pfedme_Electronics_converge}
     \end{subfigure}\hfill
     \begin{subfigure}[b]{0.235\textwidth}
         \centering
         \includegraphics[width=1\textwidth]{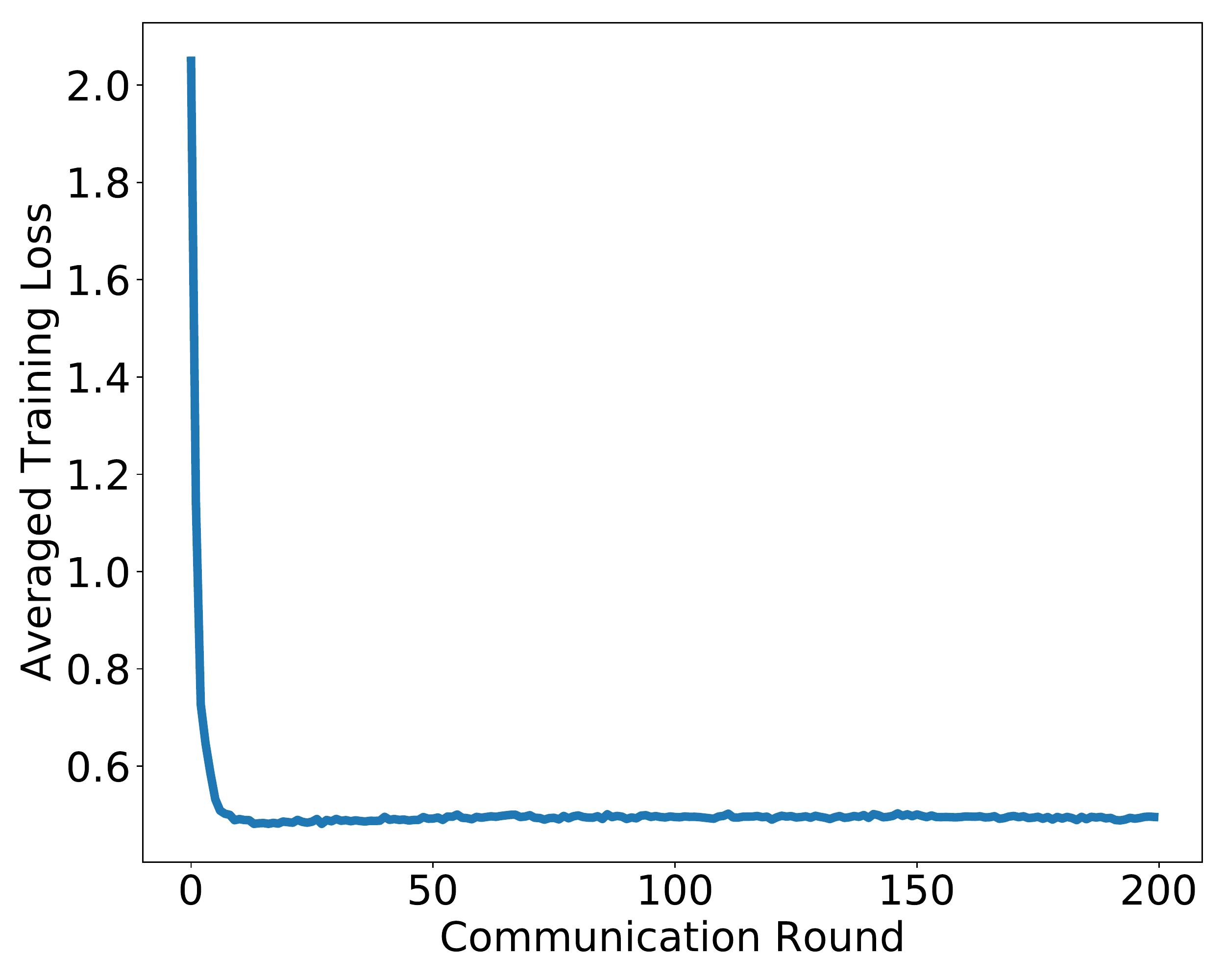}
         \caption{{\bf PF-GNN+}}
         \label{fig:pfedmestruct_Electronics_converge}
     \end{subfigure}
\caption{Empirical comparison between {\bf PF-GNN} and {\bf PF-GNN+} on the \emph{Electronics} domain.}
\label{fig:convergence}
\end{figure}
\subsection{Empirical Convergence Analysis}
\label{sec:convergence}
We show the empirical convergence analysis of both {\bf PF-GNN} and {\bf PF-GNN+} in Fig.~\ref{fig:convergence}. Both models converge as the number of global communication rounds increases, demonstrating that the bi-level optimization can minimize the loss even if heterogeneity exists across market segments. By comparing {\bf PF-GNN} and {\bf PF-GNN+}, {\bf PF-GNN} has more fluctuations than {\bf PF-GNN+}. Moreover, {\bf PF-GNN+} achieves faster convergence than {\bf PF-GNN}, which demonstrates the necessity of modeling statistical structural information in each market segment's item-item graph. 

\subsection{Sensitivity Analysis}
\label{sec:sensitivity_analysis}
For empirical thoroughness, we also investigate the influence of our proposed GNN model parameters adaptation and the graph summary adaptation on overall performance. Due to limited space, we defer this to Appendix 8. 
\begin{figure}[]
\centering
     \begin{subfigure}[b]{0.235\textwidth}
         \centering
         \includegraphics[width=1\textwidth]{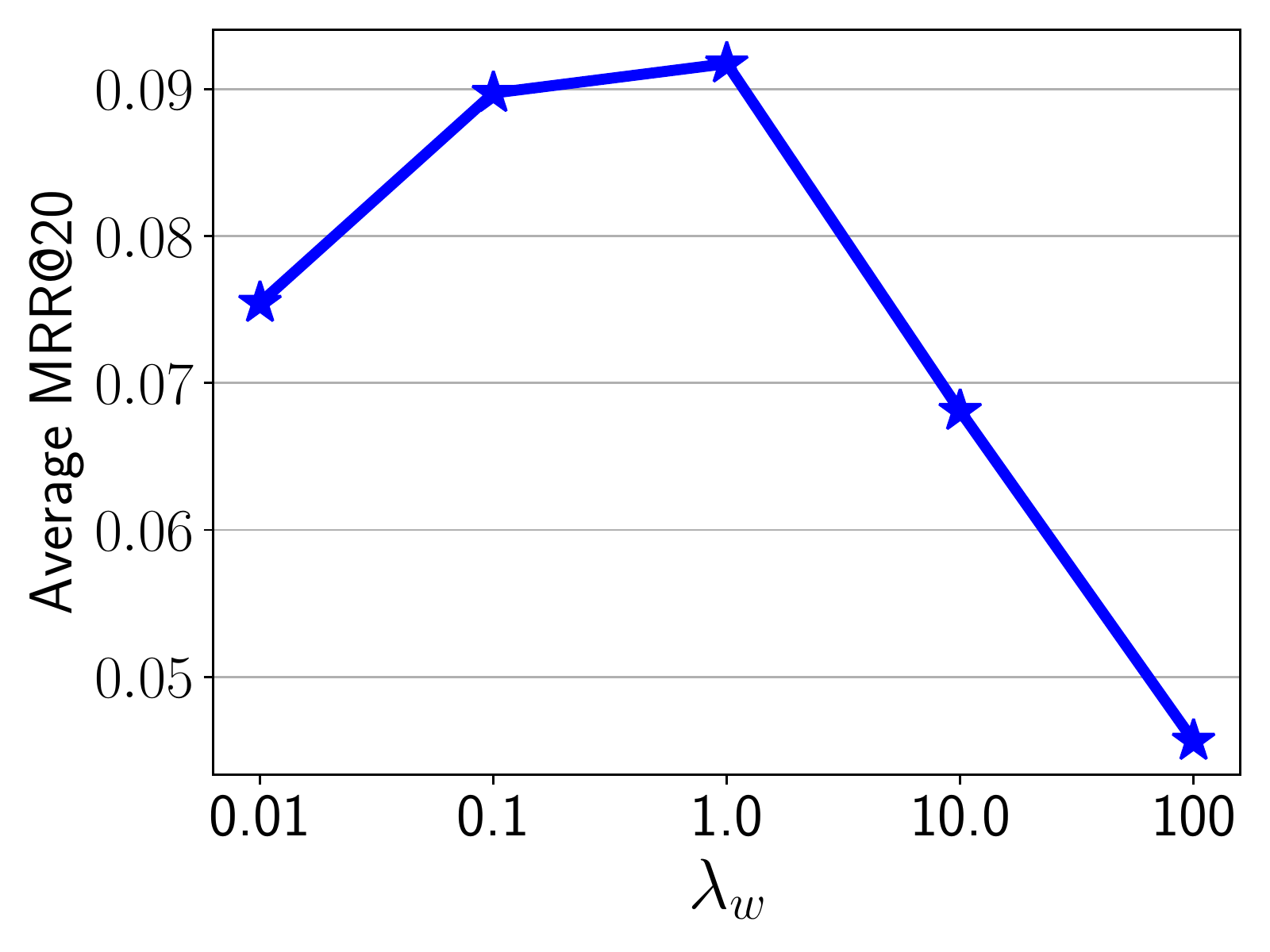}\vspace{-2mm}
         \caption{Sensitivity to $\lambda_w$}\vspace{-2mm}
         \label{fig:lambdaw_sensitivity}
     \end{subfigure}\hfill
     \begin{subfigure}[b]{0.235\textwidth}
         \centering
         \includegraphics[width=1\textwidth]{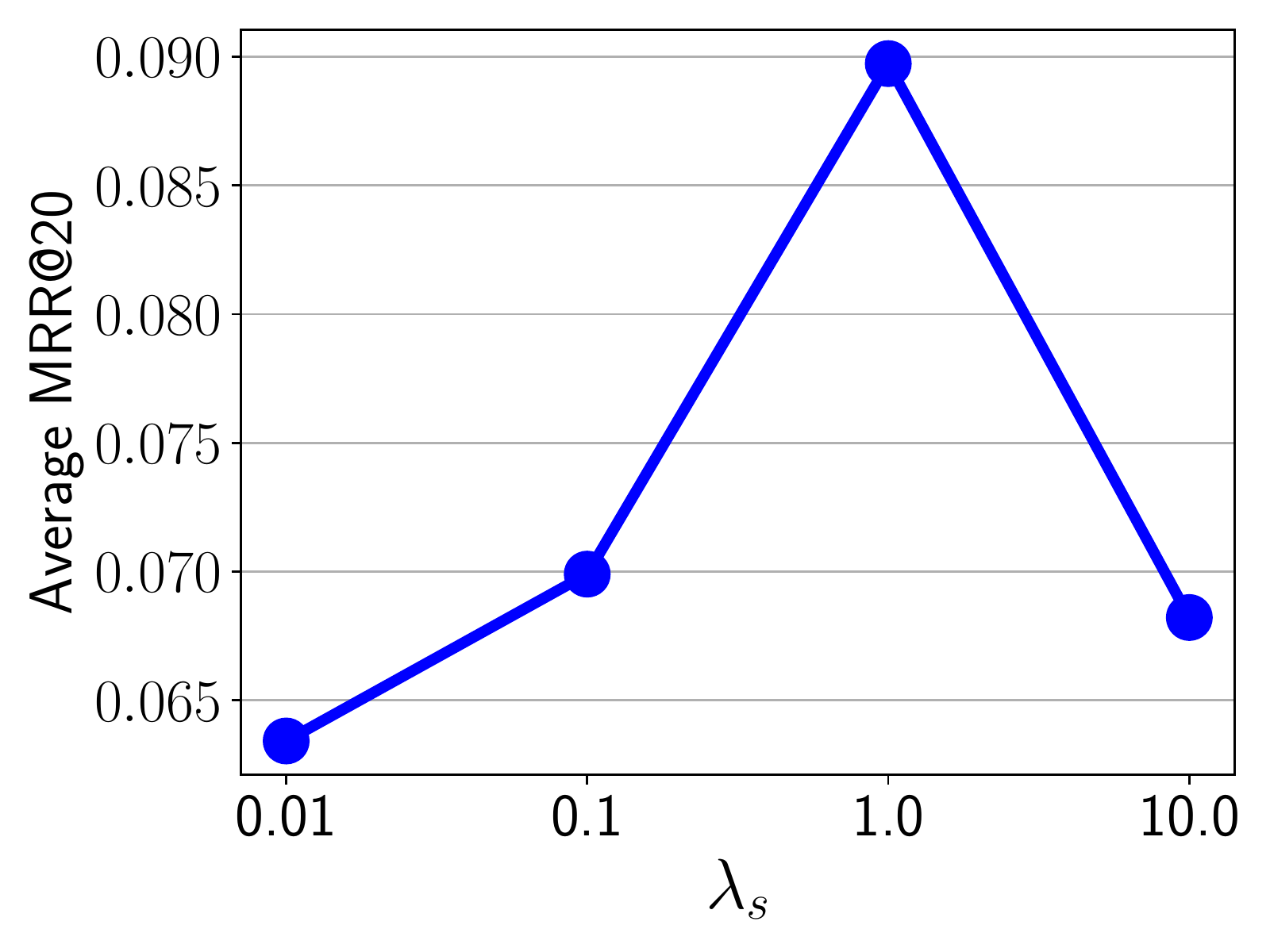}\vspace{-2mm}
         \caption{Sensitivity to $\lambda_s$}\vspace{-2mm}
         \label{fig:lambdas_sensitivity}
     \end{subfigure}
\caption{Performance Sensitivity (averaged over all markets) with respect to variation in (a) $\lambda_w$ which moderates the adaptation degree of learnable parameters $\phi$; and (b) $\lambda_s$ which regulates the adaptation degree of graph summary $\xi$.}\vspace{-6mm}
\label{fig:sensitivity}
\end{figure}




\section{Related Work}
\label{app:related}

\subsection{Item-to-Item Recommendation}
\label{sec:i2i_rec}
Item-to-item~(I2I) recommendation is a crucial component in recommender systems. I2I recommendation has several widely applied scenarios, including \emph{you may also like} in E-commerce homepages and \emph{because you watched} in video-streaming services. Existing I2I recommendation work adopts the item metadata and ID to infer the item embedding and proposes novel distance metrics for item-item affinity evaluation. One representative work is semi-parametric embedding~(SPE)~\citep{Li2019}, which adopts the mixture of ID embedding and item metadata to infer the item embedding. A pioneering work in this direction is SLIM~\citep{Karypis2011}, which proposes to model the item-item correlation weight matrix via collaborative filtering but does not account for item metadata or their higher-order interaction. Graph Neural Networks~(GNNs), which have demonstrated superiority in modeling high-order connectivity information in graph data, have been recently adopted to boost the performance of item recommendation. In fact, several GNN-based recommendation models have been proposed, which (most notably) include NGCF~\citep{wang2019neural} and LightGCN~\citep{he2020lightgcn}. However, these I2I methods assume the possibility of a centralized graph data storage which is often not practical when sharing transaction information across separate market segments is not allowed.

\subsection{Federated Learning for GNNs}
\label{sec:fed}
Federated Learning~(FL) provides new possibilities for training a global model with decentralized data privately owned by multiple clients. This is achieved via the pioneering work FedAvg of \citep{McMahan17}, which assumes all local datasets are independent and identically distributed. However, in some practical cases, this assumption is often violated when the clients collect data from heterogeneous environments. For example, in the recommendation, the item-item graphs acquired from different markets are often generated by heterogeneous preferential behaviors over a wide range of user demographics. To accommodate for this, personalized FL has been recently proposed which learns both the global model on the server and according personalized models hosted at each client node. 

Notably, Per-FedAvg~\citep{fallah2020personalized} formulates personalized FL following the model-agnostic meta-learning setup, which introduces the potential application to domain adaptation. Alternatively, pFedMe~\citep{DBLP:journals/corr/abs-2006-08848} extends FedAvg with an additional bi-level optimization regularization that moderates the deviation between each client model and the global model. Cluster FL~\citep{sattler2020clustered} proposes to apply clustering on clients so that clients in the same cluster follow similar data distribution. However, most existing personalized FL works assume a homogeneous, centralized model specification. However, this is not suitable to the context of graph-based models in item-to-item recommendation scenarios where a part of the model specification is the graph that is not fully visible to each client. In fact, each client only has access to a sub-graph of the entire item-item graph due to strict regulations concerning the storing and sharing of customer data. As such, most existing personalized FL methods cannot be applied straightforwardly to our setting.

Also, to the best of our knowledge, there have been several proposals of federated learning for GNNs with decentralized graph data in recent years~\citep{liu2022federated}, which (most notably) include GCFL+~\citep{xie2021federated} and FedGNN~\citep{wu2021fedgnn}. However, GCFL+ focuses on the graph classification task, which assumes local graphs are completed graphs instead of being fragments of a global graph (as is the case in our setting). Therefore, the proposed GCFL+ solution is applicable to GNNs that are parameterized only by the feature aggregation weights while treating the graph as the art of the input instead of part of the model specification. This does not apply to our scenario where local graphs need to be merged (without being shared explicitly) into a global graph which is part of the federated model specification. FedGNN, on the other hand, motivates the development of a federated user-centric recommender via GNN. Nonetheless, its encryption mechanism is discrete in nature and cannot be readily integrated into the gradient-based optimization framework of personalized federated learning. In our experiment, it was adapted into our baseline {~\bf F-GNN}, which ignores the encryption mechanism.

Similarly, there are also several federated learning for recommendation using local graph data from client nodes, which include (most notably) DeepRec~\citep{han2021deeprec} and MetaMF~\citep{lin2020meta}. However, like GCFL+, these works do not focus on constructing global graphs from local fragments. The proposed solutions also focus exclusively on building a common recommendation model rather than catering towards personalized models, which are specifically tailored to different user distributions that constitute different market segments. As such, these works also do not apply straightforwardly to our scenarios. With that, we believe our work on federated domain adaptation for item-item recommendation is the first that explores a potential combination between personalized FL and GNN models, which are parameterized by both (1) the graph that characterizes local interactions between feature components and (2) the combination weights that aggregate them.

\section{Conclusion}
\label{sec:conclude}
We studied the problem of personalized domain adaptation on graph neural networks with decentralized item-interaction graphs and item features across different markets. We obviate the need to centralize such data for domain adaptation by incorporating recent advances in personalized federated learning into GNNs, resulting in a personalized federated GNN framework. The proposed framework is capable of simultaneously summarizing and assembling both the structure embeddings of item-interaction graphs and the combination weights of item features across markets into a base model, which is optimized for its high adaptability towards individual markets, resulting in highly effective personalized GNN recommenders for new markets with scarce data. Our work is the first that expands on this connection between (non-graph) personalized FL and GNNs, which achieves promising results on a realistic cross-market item-to-item recommendation dataset.

\bibliography{ref}
\end{document}